\documentclass[a4paper,11pt]{article}
\pdfoutput=1 

\usepackage{jcappub,wrapfig,verbatim,amsthm,graphicx,hyperref,enumerate} 
                     
\usepackage[utf8]{inputenc}

\newcommand{\f}{f_{\mu\nu}}
\newcommand{\g}{g_{\mu\nu}}
\newcommand{\Vg}{V^\mu{}_\nu}
\newcommand{\Vf}{\widetilde{V}^\mu{}_\nu}
\newcommand{\Gg}{G^\mu{}_\nu}
\newcommand{\Gf}{\widetilde{G}^\mu{}_\nu}
\newcommand{\Tg}{T^\mu{}_\nu}
\newcommand{\Tf}{\widetilde{T}^\mu{}_\nu}
\newcommand{\Teff}{\mathcal{T}^\mu{}_\nu}
\newcommand{\fTeff}{\widetilde{\mathcal{T}}^\mu{}_\nu}

\newcommand{\wt}{\widetilde}

\newcommand{\mfphat}{m_\mathrm{FP}}
\newcommand{\omegaleff}{\Omega_\Lambda}

\newcommand{\omegade}{\Omega_\mathrm{DE}}
\newcommand{\omegadef}{\wt{\Omega}_\mathrm{DE}}
\newcommand{\meffhat}{m_\mathrm{eff}}
\newcommand{\lH}{\lambda_H}
\newcommand{\mfp}{m_\mathrm{FP}}
\newcommand{\kgeffm}{\kappa_g^\mathrm{massive}}

\newcommand{\kgeff}{\kappa_g^\mathrm{eff}}
\newcommand{\Ebr}{E_\mathrm{BR}}
\newcommand{\Egr}{E_\mathrm{GR}}

\newcommand{\omegamnot}{\Omega_{m,0}}
\newcommand{\omhat}{\widehat{\Omega}_m}

\title{\boldmath Analytical constraints on bimetric gravity}

\author{Marcus Högås, Edvard Mörtsell}

\affiliation{The Oskar Klein Centre, Department of Physics, Stockholm University, SE 106 91, Stockholm, Sweden}

\emailAdd{marcus.hogas@fysik.su.se, edvard@fysik.su.se}

\abstract{Ghost-free bimetric gravity is an extension of general relativity, featuring a massive spin-2 field coupled to gravity. We parameterize the theory with a set of observables having specific physical interpretations. For the background cosmology and the static, spherically symmetric solutions (for example approximating the gravitational potential of the solar system), there are four directions in the parameter space in which general relativity is approached. Requiring that there is a working screening mechanism and a nonsingular evolution of the Universe, we place analytical constraints on the parameter space which rule out many of the models studied in the literature. Cosmological solutions where the accelerated expansion of the Universe is explained by the dynamical interaction of the massive spin-2 field rather than by a cosmological constant, are still viable.}

\begin{document}
\maketitle
\flushbottom

\section{Introduction and summary}
There are strong motivations to look for new theories of gravity, for example the unknown nature of dark matter and dark energy. In general relativity, the accelerated expansion of the Universe can be explained by a cosmological constant $\Lambda$. To be compatible with observations we also have to include standard matter particles (e.g., baryons and photons) and cold dark matter. This is the cosmological concordance $\Lambda$CDM model. However, the model exhibits a $>4 \sigma$ discrepancy in the measured value of the Hubble constant, using information from the early and late universe \cite{Verde:2019ivm}. On the other hand, on solar system scales and below, general relativity (GR) accounts for gravitational observations to an astonishing precision \cite{Will:2014kxa}. Hence, the new theory that we are looking for should ideally exhibit new features on cosmological scales but have a screening mechanism restoring GR on smaller scales.

Bimetric gravity is a ghost-free extension of general relativity, featuring two interacting spin-2 fields, one massless (as in general relativity) and one massive. To achieve this setup, besides the physical metric $\g$, to which standard matter particles and fields couple, one must introduce a second symmetric spin-2 field (i.e., a metric) $\f$. Doing so, four constant parameters emerge, in addition to the gravitational constant and the cosmological constant which are present already in GR. This brings a rich phenomenology to the theory but also makes it more complicated to study.

Among the virtues of bimetric gravity are self-accelerating cosmological models, that is without a cosmological constant \cite{Volkov:2011an,vonStrauss:2011mq,Comelli:2011zm,Volkov:2012wp,Volkov:2012zb,Akrami:2012vf,Volkov:2013roa,Konnig:2013gxa}. An interesting consequence of these models is that a small cosmological constant is technically natural in the sense of 't Hooft which protects it against quantum corrections \cite{tHooft:1979rat}. Bimetric cosmology can push the Hubble constant in the right direction in order to ease the tension \cite{Mortsell:2018mfj}. The theory has a screening mechanism that can restore general relativity on solar system scales \cite{Babichev:2013pfa,Enander:2015kda,Platscher:2018voh}. Bimetric gravity can provide a dark matter particle \cite{Aoki:2016zgp,Babichev:2016hir,Babichev:2016bxi} and possibly get rid of some of the need of dark matter due to the screening mechanism which effectively increases the gravitational force in low density environments \cite{Enander:2015kda,Platscher:2018voh}. It should be noted that all of these features may not be achievable at the same time. What we show in this paper is that we can have self-accelerating cosmology and a working screening mechanism simultaneously whereas the results of Ref. \cite{PhysParamObs} indicate that the Hubble constant can be pushed in the right direction at the same time. 

Challenges include the presence of a gradient instability for linear perturbations around cosmological background solutions \cite{Comelli:2012db,Khosravi:2012rk,Berg:2012kn,Sakakihara:2012iq,Konnig:2014dna,Comelli:2014bqa,DeFelice:2014nja,Solomon:2014dua,Konnig:2014xva,Lagos:2014lca,Konnig:2015lfa,Aoki:2015xqa,Mortsell:2015exa,Akrami:2015qga,Hogas:2019ywm,Luben:2019yyx}. Therefore, one cannot rely on standard procedures when analyzing structure formation. Related to this is the challenge of finding a stable (well-posed) form of the equations of motion in order to obtain long-term numerical evolution of generic systems, for example of gravitational collapse \cite{Kocic:2018ddp,Kocic:2018yvr,Kocic:2019zdy,Torsello:2019tgc,Kocic:2019gxl,Torsello:2019jdg,Torsello:2019wyp,Kocic:2020pnm}. Finally, in Ref.~\cite{Hogas:2019cpg}, the question was raised whether bimetric gravity satisfies the cosmic censorship hypothesis. For reviews on bimetric gravity, see \cite{deRham:2014zqa,Schmidt-May:2015vnx}.

The bimetric action \eqref{eq:HRaction} is invariant under the constant rescaling,
\begin{equation}
\label{eq:Rescaling1}
(\f, \kappa_f , \beta_n) \to (\omega \f, \omega \kappa_f , \omega^{-n/2} \beta_n).
\end{equation}
where $\beta_n$ is the set of bimetric theory parameters (to be defined below) and $\kappa_f$ is the gravitational constant of the $\f$ metric. Hence, the $\beta$-parameters are not observables and the reported constraints on them depend on the choice of scaling, being different in different papers. The authors of Refs.~\cite{Luben:2019yyx,Luben:2020xll} proposed a parametrization for a subset of models in terms of three observables: the mixing angle $\theta$ between the massless and massive gravitons (related to $\bar{\alpha}$ in their paper), the graviton mass $\mfphat$, and the effective cosmological constant $\omegaleff$. The framework applies to models with three $\beta$-parameters or less. In this paper, we extend the physical parametrization to the general case by including two parameters $\alpha$ and $\beta$ (not to be confused with the $\beta$-parameters $\beta_0$, ..., $\beta_4$ or the $\alpha$ of Refs.~\cite{Luben:2019yyx,Luben:2020xll}) which determine how the nonlinear screening mechanism is realized. The dimensionless parameters $\Theta = (\theta , \mfphat , \omegaleff,\alpha,\beta)$ are observables that specify the theory uniquely\footnote{In the sense that they specify the free constants in the action/equations of motion. Concerning phenomenology, there can still be freedom to specify solution branches of the equation of motion.} and they have a clear physical interpretation.

\noindent Earlier results have shown that the theory is observationally viable, including cosmological data from the cosmic microwave background (CMB), baryon acoustic oscillations (BAO), and supernovae type Ia (SNIa) \cite{vonStrauss:2011mq,Akrami:2012vf,Konnig:2013gxa,Dhawan:2017leu,Lindner:2020eez,Luben:2020xll}, gravitational wave observations \cite{DeFelice:2013nba,Fasiello:2015csa,Cusin:2015pya,Max:2017flc}, solar system tests \cite{Platscher:2018voh,Luben:2018ekw}, velocity dispersion and strong gravitational lensing by galaxies \cite{Sjors:2011iv,Enander:2013kza,Enander:2015kda,Platscher:2018voh}, as well as gravitational lensing by galaxy clusters \cite{Platscher:2018voh}. However, in these works typically only a subset of observations has been used or the results apply only to a restricted set of models where one or several of the $\beta$-parameters are set to zero. Also, it is usually not clear whether the preferred values are close to any GR limit of the theory. In this paper, we treat the constraints on the theory parameters in the unified framework of the physical parameters, focusing on a set of constraints that can be expressed analytically. In the follow-up paper \cite{PhysParamObs}, we analyze the observational constraints.

In terms of the $\beta$-parameters, requiring a working screening mechanism and a consistent background cosmology, we show that one is required to include $\beta_1$, $\beta_2$, and $\beta_3$ (none of them can be zero). This implies for example that commonly studied models, for example with only $\beta_1 \beta_2$, $\beta_1 \beta_4$, and $\beta_0 \beta_1 \beta_4$ are excluded. Self-accelerating models with $\beta_0=0$ and $\beta_0 = \beta_4 = 0$ are viable. The constraints are summarized in Figs.~\ref{fig:dynhig2} and \ref{fig:ConstrSelfAcc} and a complete list of the constraints can be found in Appendix~\ref{sec:ConstraintsList}.

Within the allowed parameter space, we show that there are four limits in which the background cosmology and local solutions (for stars, galaxies, and galaxy clusters) approach GR results: small mixing angle (i.e., $\theta \to 0$) and large values of $\mfphat$, $\alpha$, or $\beta$ (i.e., $\mfphat \to \infty$, $\alpha \to \infty$, or $\beta \to \infty$). This is true for general bimetric models. Self-accelerating cosmologies are of special interest since no cosmological constant is needed to account for the accelerated expansion of the Universe. Interestingly, for these models, there is only one GR limit, $\theta \to 0$, the other three are excluded by the constraints from a consistent cosmology and requiring an existing screening mechanism.

\paragraph{Notation.} For the most part, we use geometrized units where Newton's gravitational constant and the speed of light are set to one, $G=c=1$, in which case length, time, and mass have the same units $\mathsf{L} = \mathsf{T} = \mathsf{M}$ and the mass of an object is measured in terms of half its Schwarzschild radius. Hence, the gravitational constant $\kappa_g = 8\pi G/c^4 = 8\pi$ is dimensionless.

Quantities constructed from the second metric $\f$ are denoted with tildes, otherwise constructed from the physical metric $\g$. Overdot denotes differentiation with respect to coordinate time, $\dot{} = d/dt$. Prime denotes differentiation with respect to $e$-folds, for example $\Omega_m'(a) = d \Omega_m(a) / d \ln a$.

\section{Bimetric gravity}
The Hassan--Rosen action \cite{Hassan:2011zd},
\begin{align}
\label{eq:HRaction}
\mathcal{S}_\mathrm{HR} = \int d^4 x &\big[\frac{1}{2\kappa_g}  \sqrt{- \det g} R + \frac{1}{2\kappa_f} \sqrt{- \det f} \widetilde{R} - \sqrt{-\det g} \sum_{n=0}^{4} \beta_n e_n(S) + \nonumber\\
&+ \sqrt{-\det g} \mathcal{L}_m + \sqrt{-\det f} \widetilde{\mathcal{L}}_m\big],
\end{align}
is constructed to avoid the Boulware--Deser ghost which is present in general theories of massive gravity \cite{Boulware:1973my}. $R$ is the Ricci scalar of $\g$, $\wt{R}$ is the Ricci scalar of $\f$, and $e_n(S)$ are the elementary symmetric polynomials of the square root of the two metrics, defined by $S^\mu{}_\rho S^\rho{}_\nu = g^{\mu\rho} f_{\rho\nu}$ \cite{Hassan:2017ugh,higham2008functions}. There are five $\beta$-parameters with dimension of curvature $1/\mathsf{L}^2$. For the theory to exhibit new features on cosmological scales, the curvature scale should typically be set by the Hubble radius, that is $\beta_n \sim 1/ H_0^{-2}$. At the level of the action, $\beta_0$ and $\beta_4$ are cosmological constant terms in the $g$- and $f$-sector, respectively. However, writing out the cosmological equations of motion, there appears additional (cosmological) constant terms which is due to the massive spin-2 field, as we will see below. There can be two independent matter sectors, one coupled to $\g$ and one coupled to $\f$, described by the Lagrangians $\mathcal{L}_m$ and $\wt{\mathcal{L}}_m$, respectively. The former is coupled to the physical metric and contains the standard model particles and fields (including dark matter) and the latter is coupled to the second metric and can include an independent sector of particles and fields \cite{deRham:2014naa,deRham:2014fha}.

Varying the action \eqref{eq:HRaction} with respect to $\g$ and $\f$, we derive the equations of motion,
\begin{subequations}
	\label{eq:EoM}
	\begin{alignat}{2}
	\Gg &= \kappa_g \Teff,& \qquad \Teff &\equiv \Tg + \Vg,\\
	\Gf &= \kappa_f \fTeff,& \qquad \fTeff &\equiv \Tf +  \Vf,
	\end{alignat}
\end{subequations}
where $\Gg$ and $\Gf$ are the Einstein tensors of $\g$ and $\f$, respectively. The matter stress--energies $\Tg$ and $\Tf$ are obtained by varying $\mathcal{L}_m$ and $\wt{\mathcal{L}}_m$ with respect to the metrics. From now on, we assume that there is only one matter sector, coupled to $\g$, and thus $\Tf=0$. Since this matter is coupled to $\g$, this is the metric determining the geodesics of freely falling observers. Therefore, an observer measures the geometry of $\g$ and we refer to it as the physical metric. The ratio of the gravitational constants of the two metrics, $\kappa_g$ and $\kappa_f$, is denoted,
\begin{equation}
	\kappa \equiv \kappa_g / \kappa_f.
\end{equation}
In natural units (i.e., $c=\hbar =1$), one typically refers to the reduced Planck mass instead of the gravitational constant, being related as $M_\mathrm{Pl}^2 = \hbar / c^3 \kappa_g$. The bimetric stress--energies $\Vg$ and $\Vf$ originate in the interaction between the metrics and read,
\begin{subequations}
	\label{eq:BRpots}
	\begin{align}
	\label{eq:Vg}
	\Vg &\equiv - \sum_{n=0}^{3} \beta_n \sum_{k=0}^{n} (-1)^{n+k} e_k(S) {(S^{n-k})}^\mu{}_\nu,\\
	\Vf &\equiv - \sum_{n=0}^{3}\beta_{4-n} \sum_{k=0}^{n} (-1)^{n+k} e_k(S^{-1}) {(S^{-n+k})}^\mu{}_\nu.
	\end{align}
\end{subequations}
Assuming conservation of matter stress--energy, $\nabla_\mu \Tg=0$, it follows that,
\begin{equation}
	\nabla_\mu \Vg=0.
\end{equation}

\section{Physical parameterization}
The Hassan--Rosen action \eqref{eq:HRaction} is invariant under the constant rescaling \eqref{eq:Rescaling1}. Thus, all physical quantities must be independent on the choice of $\omega$. Proportional solutions $\g = c^2 \f$ are of special interest since they describe the asymptotic structure of the static, spherically symmetric solutions as well as the future infinity of bimetric cosmology. In the physical parameterization  that we are introducing, we assume that space-time is proportional asymptotically (in time or space).\footnote{Note that the graviton mass, and hence also the mixing angle between the gravitons, can be defined only around proportional backgrounds \cite{Hassan:2012wr}.} For proportional solutions, the symmetry transformation reads,
\begin{equation}
\label{eq:Rescaling2}
(c^2 , \kappa , \beta_n) \to (\omega c^2 , \omega^{-1} \kappa, \omega^{-n/2} \beta_n).
\end{equation}
In the literature it is common to use this invariance to set $c=1$ (here the conformal factor, not to be confused with the speed of light) or $\kappa = \kappa_g / \kappa_f=1$. Instead, we define variables which are independent of this choice. Note that $\beta_n c^n$ is invariant under \eqref{eq:Rescaling2}. As a rescaling invariant alternative to the $\beta$-parameters one can introduce the dimensionless parameters,
\begin{equation}
\label{eq:Bdef}
B_n \equiv \kappa_g \beta_n c^n/H_0^2,
\end{equation}
corresponding to the $\beta$-parameters in units of the curvature scale defined by the Hubble constant. This is especially convenient for cosmological applications since the regime $\beta_n \sim H_0^2$ is where we expect to find a modified background cosmology, that is when $B_n \sim 1$ (cf., \cite{Mortsell:2017fog}).

In the action \eqref{eq:HRaction}, seven constant parameters appear: $\kappa_g$, $\kappa_f$, and $\beta_0$, ..., $\beta_4$. However, $\kappa_g = 8\pi$. Due to the rescaling invariance, there is one redundant parameter. Hence, there are five independent parameters in bimetric theory. Instead of $(\kappa_f,\beta_0 , ..., \beta_4)$ with one of them being redundant, one can parameterize the bimetric interaction in terms of the dimensionless parameters $\mathrm{\Theta} \equiv (\theta,\mfphat,\omegaleff,\alpha,\beta)$ which are independent of the rescaling. In terms of the $B$-parameters, they are given by,
\begin{subequations}
	\label{eq:PhysToBeta}
	\begin{align}
		\label{eq:ThetaInB}
		\tan^2 \theta &= \frac{B_1 + 3 B_2 + 3 B_3 + B_4}{B_0 + 3 B_1 + 3 B_2 + B_3},\\
		\mfphat^2 &= \left(B_1 + 2 B_2+ B_3\right) / \sin^2 \theta,\\
		\omegaleff &= \frac{B_0}{3} + B_1 + B_2 + \frac{B_3}{3},\\
		\alpha &= - \frac{B_2 + B_3}{B_1 + 2 B_2 + B_3},\\
		\beta &=\frac{B_3}{B_1 + 2 B_2 + B_3}.
	\end{align}
\end{subequations}
Here, $\theta$ is the mixing angle between the massless and massive gravitons, $\mfphat$ is the mass of the graviton as measured in units of $H_0 \sim 10^{-33} \, \mathrm{eV}/c^2$, $\omegaleff$ is the effective cosmological constant, and $\alpha$ and $\beta$ are parameters that determine the screening mechanism. We will frequently refer to the Compton wavelength $\lambda_g = 1 / (H_0 \mfphat)$ instead of $\mfphat$ reflecting the same physical parameter. The equations \eqref{eq:PhysToBeta} can be inverted with the result,
\begin{subequations}
	\label{eq:BetaToPhys}
	\begin{align}
	\label{eq:B0phys}
	B_0 &= 3 \omegaleff - \sin^2 \theta \, \mfphat^2 (3 + 3\alpha + \beta),\\
	B_1 &= \sin^2 \theta \, \mfphat^2 (1+ 2 \alpha + \beta),\\
	B_2 &= - \sin^2 \theta \, \mfphat^2 (\alpha + \beta),\\
	B_3 &= \sin^2 \theta \, \mfphat^2 \beta,\\
	B_4 &= 3 \tan^2 \theta \, \omegaleff + \sin^2 \theta \, \mfphat^2 (-1 + \alpha - \beta).
	\end{align}
\end{subequations}
Note that $B_1,B_2,B_3,B_4 \to 0$ in the limit $\theta \to 0$. Hence, the bimetric stress--energy \eqref{eq:Vg} contributes with a cosmological constant term in the equations of motion for the physical metric and the theory reduces to general relativity.\footnote{For a discussion of the strong-coupling scale of the second metric in this limit, see Ref. \cite{Akrami:2015qga}.} The Jacobian determinant of \eqref{eq:BetaToPhys} is $18 \sin^6 \theta \, \mfphat^5 \omegaleff $ and thus there is a one-to-one relation between the $B$-parameters and the physical parameters, except from the special cases where at least one of $\theta$, $\mfphat$, or $\omegaleff$ vanishes (the values of $\alpha$ and $\beta$ do not affect the Jacobian determinant). Thus, the physical parameters determine uniquely the free constants in the action \eqref{eq:HRaction}. However, when there is such a choice, there is still freedom to select different solution branches of the equations of motion, see Section~\ref{sec:Cosmo} for example. Also, we defined the physical parameters making use of the proportionality constant $c$ in the definition of $B_n$ \eqref{eq:Bdef}. Hence, we have implicitly assumed that the solutions we are interested in are proportional asymptotically (in time or space).

\section{Local solutions}
Static, spherically symmetric (SSS) solutions can be used to approximate the gravitational potential of for example the solar system and galaxies. We refer to them as local solutions, as opposed to cosmological solutions. In general relativity, Birkhoff’s theorem states that the Schwarzschild metric is the unique spherically symmetric vacuum solution \cite{Jebsen:1921,Jebsen2005,Birkhoff:1923}. In bimetric relativity, there is no such theorem \cite{Kocic:2017hve}. There are two types of approximate analytical solutions to the SSS equations of motion in bimetric gravity, applicable in two different regimes: the linearized solutions and the nonlinear Vainshtein solutions to which the following two sections are devoted.

\subsection{Linearized solutions}
Linearizing around a proportional background $\f = c^2 \g$ with constant $c$, we get one massless gravitational mode ($\delta G_{\mu\nu}$, not to be confused with the Einstein tensor) and one massive gravitational mode  ($\delta M_{\mu\nu}$) \cite{Hassan:2012wr},
\begin{equation}
\label{eq:Linearized}
	\left(\begin{array}{c}
	\delta G_{\mu\nu}\\
	\delta M_{\mu\nu}
	\end{array}\right) = \left(\begin{array}{rr}
	\cos \theta & \sin \theta \\
	-\sin \theta & \cos \theta
	\end{array}\right) \left(\begin{array}{c}
	\delta \g\\
	\delta \f
	\end{array}\right),
\end{equation}
where the angle $0 \leq \theta \leq \pi / 2$ is defined by,
\begin{equation}
\label{eq:kbarDef}
	\tan^2 \theta = \kappa c^2.
\end{equation}
The matrix in \eqref{eq:Linearized} is a rotation matrix, so $\theta$ is the mixing angle of the massive and massless gravitons. Note that $\kappa c^2$ and hence $\theta$ are independent of the rescaling invariance. In the limit $\theta \to 0$, the physical metric coincides with the massless spin-2 field and we recover GR. In the limit $\theta \to \pi / 2$, the physical metric coincides with the massive spin-2 field and we recover dRGT (de Rham--Gabadadze--Tolley) massive gravity \cite{deRham:2010ik,deRham:2010kj} with a fixed second metric, see for example \cite{Hassan:2012wr,Hassan:2014vja}. These two limits hold not only for the linear perturbations around proportional backgrounds but also nonlinearly, which can be seen from the action \eqref{eq:HRaction}.

For proportional metrics which are homogeneous and isotropic (flat Minkowski space-time being one example), the equations of motion require the effective cosmological constants of the two metric sectors to be equal, corresponding to,
\begin{equation}
\label{eq:OmegagOmegaf}
\frac{B_0}{3} + B_1 + B_2 + \frac{B_3}{3}= \frac{1}{\tan^2 \theta} \left(\frac{B_1}{3} + B_2 + B_3 + \frac{B_4}{3}\right),
\end{equation}
see also \eqref{eq:OmegaDefs} and \eqref{eq:BRFriedm_Alt}. This equation allows us to express $\theta$ in terms of the $B$-parameters as in \eqref{eq:ThetaInB}. In the presence of a source of mass $M$ (e.g., a star or a galaxy), the linearized gravitational potential $\Phi$ and scalar curvature $\Psi$ acquire exponential Yukawa terms in addition to the Newtonian $1/r$ potential \cite{Comelli:2011wq},
\begin{subequations}
	\label{eq:LinPot}
	\begin{align}
		\Phi(r) &= - M \cos^2 \theta \left(\frac{1}{r} + \frac{4\tan^2 \theta}{3} \frac{e^{-r/\lambda_g}}{r}\right),\\
		\Psi(r) &= - M \cos^2 \theta \left(\frac{1}{r} + \frac{2 \tan^2 \theta (1+r/\lambda_g)}{3} \frac{e^{- r/\lambda_g}}{r}\right).
	\end{align}
\end{subequations}
In the dRGT massive gravity limit ($\theta \to \pi/2$), only the Yukawa term survives. The solution \eqref{eq:LinPot} is obtained assuming a flat background. A perhaps more appropriate background for our universe would be de Sitter with a cosmological constant $\Lambda$. However, if $\Lambda \sim H_0^2 \equiv 1/\lH^2$, the effect of including the cosmological constant becomes visible only at length scales of the order of the Hubble radius $\lH$ which is much greater than the size any local system anyway.

In \eqref{eq:LinPot}, $\lambda_g$ is the Compton wavelength of the massive graviton $\lambda_g = 1 / (H_0 \mfphat)$, and the graviton mass (Fierz--Pauli mass) is given by,
\begin{equation}
\label{eq:mFPhat}
\mfphat^2 = \left(B_1 + 2 B_2+ B_3\right) / \sin^2 \theta.
\end{equation}
Note that $\mfphat$ is independent of the rescaling invariance. Far outside the Compton wavelength $r\gg \lambda_g$, the gravitational potentials reduce to general relativity with a gravitational constant $\kappa_g \cos^2 \theta$ as can be seen from \eqref{eq:LinPot}. However, viable background cosmologies are usually obtained with a $\lambda_g$ which is of the same order of magnitude as the Hubble radius, $\lambda_g \sim \lH$. In this case, the $r \gg \lambda_g$ regime is observationally irrelevant. 

As we see in \eqref{eq:LinPot}, due to the exponential Yukawa terms, the linearized solutions deviate from general relativity. That is, unless $\theta \to 0$ or $\lambda_g \to 0$ (i.e., $\mfphat \to \infty$) in which case the Yukawa terms are suppressed. Naively, one would expect to recover GR also in the zero mass limit $\mfphat \to 0$. However, in this limit the gravitational slip for the local system $\gamma \equiv \Phi/\varphi$ does not approach the GR result, which is unity, but,
\begin{equation}
	\gamma|_{\mfphat \to 0} = [5 + \cos (2\theta)]/6.
\end{equation}
Here, $\varphi$ is the Weyl potential $\varphi \equiv (\Phi + \Psi)/2$. Unless $\theta =0$, the gravitational slip $\gamma < 1$ in the zero mass limit, meaning that the bending of light around SSS sources will be different from GR. This is a vDVZ-like (van Dam--Veltman--Zakharov) discontinuity \cite{vanDam:1970vg,Zakharov:1970cc}. 

Since GR describes gravitational observations to a very high accuracy on local scales \cite{Will:2014kxa}, the linearized solution cannot be applicable on these scales unless $\theta$ or $\lambda_g$ are small \cite{Luben:2018ekw}. However, in the limit $\theta \to 0$, general relativity is approached and in the $\lambda_g \to 0$ limit (i.e., $\mfphat \to \infty$), general relativity is approached at the level of the cosmological background solutions (assuming that we impose no background independent value on the matter density, see Section~\ref{sec:GRlims}). In both these cases many of the interesting features of the theory disappear. To ensure the existence of novel cosmological solutions while at the same time satisfying solar system tests, we demand the existence of a screening mechanism hiding the additional bimetric degrees of freedom in for example the solar system. It should be noted that this requirement is a conservative one and that there may be novel cosmological solutions satisfying for example solar system tests via suppression of the Yukawa term while still not being completely in the $\theta \to 0$ or $\lambda_g \to 0$ limits. A screening mechanism for massive gravity was first proposed by Vainshtein, suggesting that the higher order terms become important and restore general relativity at the nonlinear level, thereby invalidating the linear solution \cite{Vainshtein:1972sx}.

\subsection{Vainsthein screening}
In bimetric gravity, the linearized solution that we presented in the previous section is only valid above a certain length scale, usually taken to be the Vainsthein radius $r_V$ (see below). If $r \lesssim r_V$, higher order terms become of the same order of magnitude as the linear terms and cannot be ignored \cite{Enander:2013kza,Babichev:2013pfa}.\footnote{In Ref.~\cite{Enander:2013kza}, the Vainshtein radius is defined as the radius where the second order terms become of the same order of magnitude as the first order ones. In this paper, we solve the full nonlinear equations while adopting \eqref{eq:VainshteinDef} as the definition of the Vainshtein radius.} These nonlinear terms can effectively restore general relativity within some radius. 

\begin{figure}[t]
	\centering
	\includegraphics[width=\linewidth]{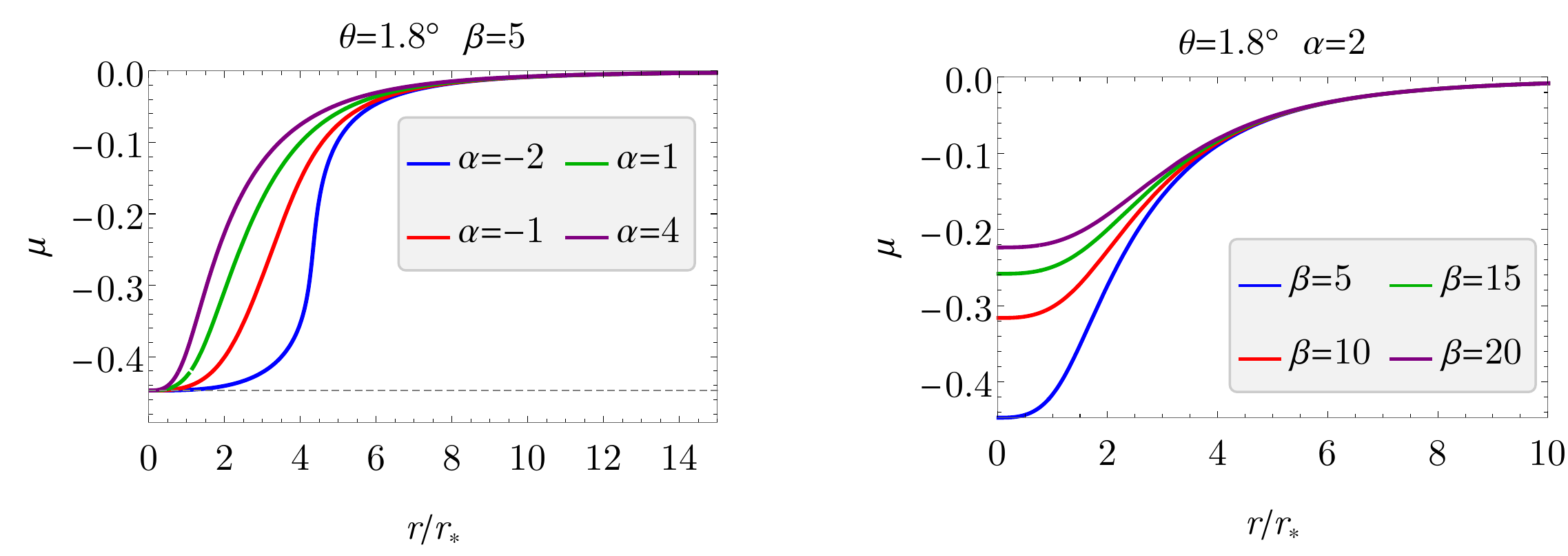}
	\caption{Examples of $\mu(r)$. Here, as an arbitrarily chosen example, we have set $\theta = 1.8^{\circ}$. The Stückelberg field $\mu(r)$ has two plateaus, one as $r \to 0$ and the other as $r \to \infty$. Increasing the value of $\alpha$ gives an earlier transition between the plateaus and in the limit $\alpha \to \infty$, $\mu(r) \to 0$. Increasing $\beta$ brings the $\mu(r=0)$ value closer to zero and in the limit $\beta \to \infty$, $\mu(r) \to 0$. The value at $r=0$ is set by $-1/\sqrt{\beta}$ as indicated by the gray dashed line in the left panel. At $r \gg r_V$, $\mu(r) \propto - (r_V / r)^3$. For these solutions we are assuming that we are outside of the source. When approaching the Schwarzschild radius $r \to r_S$, the weak-field approximation breaks down. }
	\label{fig:mu-example}
\end{figure}

As shown in Refs.~\cite{Babichev:2013pfa,Enander:2015kda}, if we are outside a source of radius $r_*$ (i.e., where the pressure is vanishing $P=0$ and $M = \mathrm{const.}$) yet far inside the Compton wavelength $\lambda_g$, the equations for the gravitational potentials can be written in terms of a Stückelberg field $\mu(r)$ (see Appendix~\ref{sec:muEq} for an explanation of that name),
\begin{subequations}
	\label{eq:VainsteinPots}
	\begin{alignat}{2}
		r \partial_r \Phi(r) &= +\frac{2M}{r} - \sin^2 \theta \left(\frac{r}{\lambda_g}\right)^2 \left[\mu(r) - \frac{\beta}{3}\mu^3(r)\right],& \quad r &> r_*,\\
		\Psi(r) &= - \frac{2M}{r} - \sin^2 \theta \left(\frac{r}{\lambda_g}\right)^2 \left[\mu(r) - \alpha \mu^2(r) + \frac{\beta}{3} \mu^3(r)\right],& \quad r &> r_*,
	\end{alignat}
\end{subequations}
where,
\begin{equation}
	\label{eq:AlphaBeta}
	\alpha \equiv - \frac{B_2 + B_3}{B_1 + 2 B_2 + B_3}, \quad 
	\beta \equiv \frac{B_3}{B_1 + 2 B_2 + B_3},	
\end{equation}
and $\mu(r)$ is the solution to a polynomial of degree seven, see Appendix~\ref{sec:muEq}. The physical parameters $\alpha$ and $\beta$ are independent of the rescaling invariance and determine the shape of $\mu(r)$ and hence also the gravitational potentials, see Fig.~\ref{fig:mu-example} for some typical examples. Here, we have assumed that we are outside the matter source. Inside the source, we may also have to take the pressure $P(r)$ into account as well as the radial dependence of the mass $M(r)$. Inside a homogeneous source where the matter density and pressure are constant, $\mu(r)$ is constant.

Outside the source, $\lambda_g$, $M$, and $r$ appears only in the combination $M \lambda_g^2 /r^3$ in the equation defining $\mu$ \eqref{eq:muPoly}. Therefore, the radius within which GR can be restored is proportional to $\left(r_S \lambda_g^2 \right)^{1/3}$ where $r_S = 2M$ is the Schwarzschild radius of the source. The equation which determines $\mu(r)$, \eqref{eq:muPoly}, also contains $\theta$, $\alpha$, and $\beta$. Hence, the radius within which GR can be restored depends also on these parameters although the dependence is more complicated than for $r_S$ and $\lambda_g$.

For simplicity, we adopt,
\begin{equation}
\label{eq:VainshteinDef}
r_V \equiv \left(r_S \lambda_g^2 \right)^{1/3},
\end{equation}
as the definition of $r_V$, keeping in mind that only if $\alpha \sim \beta \sim 1$, it represents the radius within which we start to approach GR. If we push $\alpha$ or $\beta$ away from unity, this radius increases, see Fig.~\ref{fig:incralphabeta}.

As an example, to have a viable background cosmology, typically $\lambda_g \sim \lH$ and the Vainshtein radius,
\begin{equation}
	r_V / \lambda_g \sim r_V / \lH \sim \left(r_S / \lH\right)^{1/3} \ll 1, \quad \lambda_g \sim \lH.
\end{equation}
Thus, the Vainshtein radius lies far inside the Compton wavelength of the massive graviton. Moreover, in the case $\lambda_g \sim \lH$,
\begin{equation}
	r_V / r_* = \left(\rho_* / \rho_c \mfphat^2\right)^{1/3} \sim \left(\rho_* / \rho_c\right)^{1/3}, \quad \lambda_g \sim \lH,
\end{equation}
where $\rho_* = M_* / (4\pi r_*^3/3)$ is the mean density. In this case, any object which has a mean density greater than the critical density $\rho_c \equiv 3H_0^2 / \kappa_g$ of the Universe, the Vainshtein radius will lie outside the radius of the source.\footnote{If the energy density today, including standard matter (pressureless dust and radiation) and dark energy, is equal to the critical density, then the spatial geometry of the Universe is flat (i.e., Euclidean).} This includes the solar system, galaxies, and galaxy clusters, for which the Vainshtein mechanism is active. In this case, for the Sun the Vainshtein radius lies far outside the solar system at $r_V \sim 10^7 \,  \mathrm{AU}$ where the gravitational force is negligible anyway. For galactic objects, $r_V$ is closer to the radius of the galaxy. For example, for the Milky way, $r_V^\mathrm{Milky \, Way} \sim 100 \, r_*$ where we used $M \sim 10^{12} M_\odot$ and $r_* \sim 50 \, 000 \, \mathrm{ly}$.

We can define an effective distance dependent gravitational ``constant" experienced by massive (non-relativistic) particles,
\begin{equation}
\kgeffm(r) \equiv  \kappa_g \Phi(r)/\Phi_\mathrm{GR}(r), \quad \Phi_\mathrm{GR}(r) = -2M/r.
\end{equation}
From \eqref{eq:VainsteinPots}, we see that in the two limits $r \ll r_V$ and $r \gg r_V$, $\kgeffm$ approaches,
\begin{subequations}
	\begin{alignat}{2}
		\kgeffm &= \kappa_g,& \quad r &\ll r_V,\\
		\kgeffm &= \kappa_g\left(1+ \frac{1}{3}\sin^2 \theta\right),& \quad r &\gg r_V.
	\end{alignat}
\end{subequations}
Hence, the effective gravitational force increases further out from the center with the result that less dark matter is needed in order to flatten out the galaxy rotation curves, see Refs.~\cite{Enander:2015kda,Platscher:2018voh}. The value at $r \gg r_V$ is set by $\theta$. If we move even further out, beyond the Compton wavelength of the massive graviton, the linearized solution is valid and as $r \to \infty$ we approach GR with an effective gravitational constant for massive particles,
\begin{equation}
\label{eq:kappagMassive}
	\kappa_g^\mathrm{massive} = \kappa_g \cos^2 \theta, \quad r \gg \lambda_g.
\end{equation}
\begin{figure}[t]
	\centering
	\includegraphics[width=\linewidth]{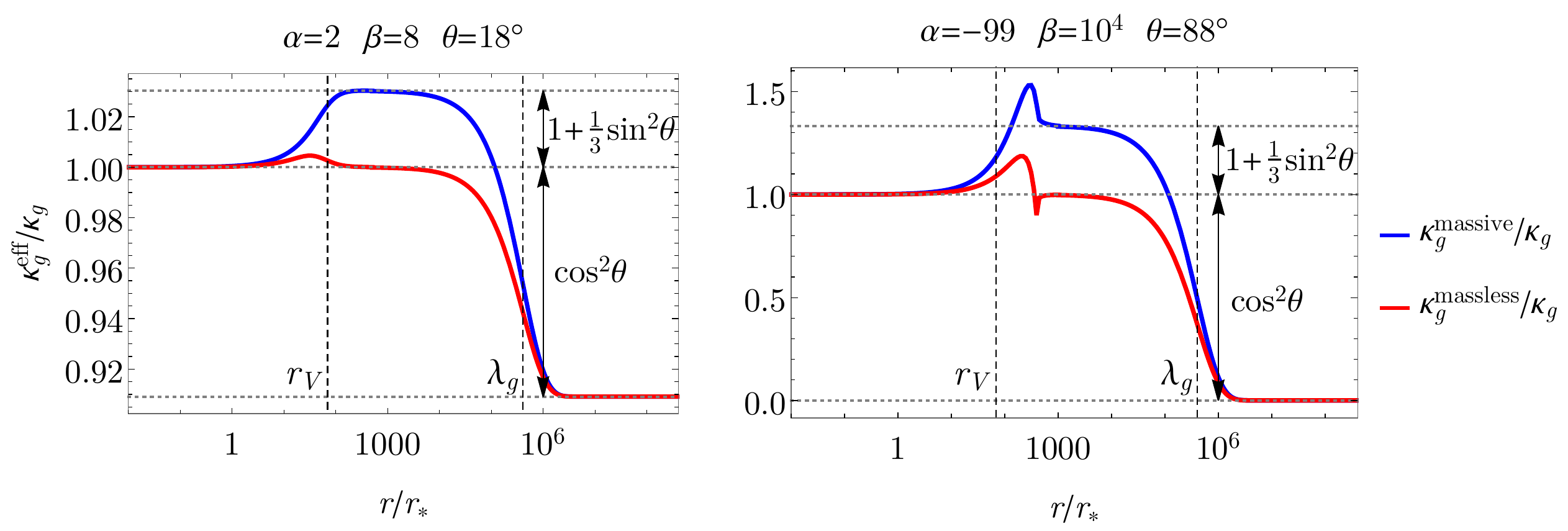}
	\caption{Effective gravitational ``constants" for massive (blue) and masseless (red) particles. The two panels show two different choices of $\alpha$, $\beta$, and $\theta$. The vertical dashed lines indicate $r_V$ and $\lambda_g$. We have set the Compton wavelength to $\lambda_g = \lH = 4 \times 10^{5} \, r_*$. The numbers are consistent with a galaxy with Vainshtein radius $r_V = 100 \, r_*$ and mass $M = 1.0 \times 10^{-6} \, r_*$. Well inside $\lambda_g$, the nonlinear Vainshtein solution is valid. Well outside $r_V$, the linear solution is applicable. Note the different scales on the vertical axes of the two panels.}
	\label{fig:effective-kappag}
\end{figure}
The potential for massless (relativistic) particles is \cite{Bertschinger:2011kk},
\begin{equation}
	\varphi(r) = (\Phi(r)+\Psi(r))/2,
\end{equation}
and the effective gravitational ``constant" for the massless particles can be defined as,
\begin{equation}
	\kappa_g^\mathrm{massless}(r) \equiv \kappa_g \varphi(r) / \varphi_\mathrm{GR}(r).
\end{equation}
Again, we can identify three different regions,
\begin{subequations}
	\begin{alignat}{2}
	\label{eq:kappageff_inside}
	\kappa_g^\mathrm{massless} &= \kappa_g,& \quad r &\ll r_V,\\
	\label{eq:kappageff_outside}
	\kappa_g^\mathrm{massless} &= \kappa_g,& \quad r &\gg r_V,\\
	\kappa_g^\mathrm{massless} &= \kappa_g \cos^2 \theta,& \quad r &\gg \lambda_g.
	\end{alignat}
\end{subequations}
In Fig.~\ref{fig:effective-kappag}, we plot the effective gravitational constants for two different examples.  

For negative $\alpha$, we observe an interesting phenomenon. The gravitational constant for massive particles has a peak before the $r \gg r_V$ plateau, see  Fig.~\ref{fig:effective-kappag}. The peak is most prominent in the limit $\alpha \to - \sqrt{\beta}$, which is the borderline case where we have a working Vainshtein mechanism, see Fig.~\ref{fig:vainshtein}. For a large mixing angle (i.e., close to $\theta = \pi / 2$), the peak can obtain values of at least $\kgeffm / \kappa_g \simeq 1.6$ which means that gravity is $\simeq 60 \, \%$ stronger at this radius than far inside the Vainshtein radius. This should be compared to the maximum value of the plateau which is $\kgeffm / \kappa_g \simeq 1.3$ (which is realized in when $\theta = \pi / 2$).

\section{Background cosmology}
\label{sec:Cosmo}
We imagine that the local solutions are defined at the final de Sitter (dS) point of the cosmological background solutions. At this point, the metrics are proportional $\f = c^2 \g$. In other words, the local metrics should tend to the proportional dS solution as $r \to \infty$. This allows us to identify the conformal factor $c$ of the cosmological de Sitter solution with the conformal factor $c$ of the SSS solutions.

\begin{figure}[t]
	\centering
	\includegraphics[width=\linewidth]{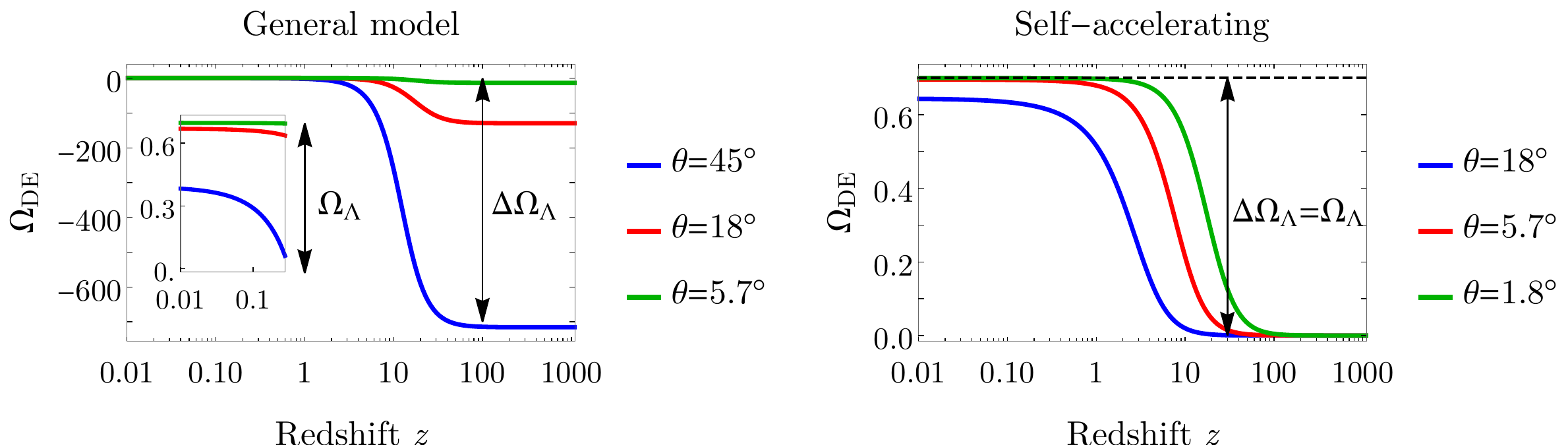}
	\caption{The dark energy density as a function of redshift. \emph{Left panel}: general models with $(\mfphat,\omegaleff,\alpha,\beta)=(10,0.7,10,10)$, examples of decreasing $\theta$. \emph{Right panel}: self-accelerating models with $(\mfphat,\omegaleff,\alpha)=(1.2,0.7,1)$, examples of decreasing $\theta$; $\beta$ can be obtained from $B_0=0$ with \eqref{eq:BetaToPhys}. For general models, the early-time dark energy density can be negative and the difference between the late-time and early-time cosmological constants is given by $\Delta \Omega_\Lambda$, according to \eqref{eq:DeltaOmegaL}. (In the left panel, we indicate $\Delta \Omega_\Lambda$ for the $\theta = 45^{\circ}$ model.) For self-accelerating models, the early-time cosmological constant vanishes. The late-time cosmological constant is set by $\omegaleff$.}
	\label{fig:omegade}
\end{figure}

\noindent Assuming that the Universe is homogeneous and isotropic with respect to both metrics in the same coordinates \cite{Torsello:2017ouh}, viable background cosmology requires the metrics to take the form,
\begin{equation}
	ds_g^2 = -dt^2 + a^2 \left(\frac{dr^2}{1-kr^2} + r^2 d\Omega^2\right), \quad ds_f^2 = - \frac{\dot{\wt{a}}^2}{\dot{a}^2} dt^2 + \wt{a}^2 \left(\frac{dr^2}{1-kr^2} + r^2 d\Omega^2\right),
\end{equation}
where $a$ and $\wt{a}$ depend on time only and $d\Omega^2$ is the metric on the unit 2-sphere ($d\Omega^2 = d\theta^2 + \sin^2 \theta d\phi^2$) and $k=$const. \cite{vonStrauss:2011mq}. The modified Friedmann equation reads (omitting the argument $t$),
\begin{equation}
\label{eq:BRFriedm}
	E^2 = \Omega_m + \Omega_k + \omegade, \quad E \equiv H / H_0.
\end{equation}
Here, $H = \dot{a}/a$ is the Hubble parameter and $H_0$ is the Hubble parameter evaluated today (i.e., the Hubble constant). $\Omega_m$ and $\Omega_k$ are dimensionless contributions from matter and spatial curvature, respectively,
\begin{equation}
		\Omega_m = \frac{\kappa_g \rho_m}{3 H_0^2} = \Omega_{m,0} (1+z)^{3(1+w_m)}, \quad \Omega_k = - \frac{k}{H_0^2 a^2} = \Omega_{k,0} (1+z)^2,
\end{equation}
with $\rho_m$ being the matter energy density. Eq. \eqref{eq:BRFriedm} can be generalized by adding several types of matter fields, but since we are mostly concerned with redshifts in the range $0 \leq z \lesssim 1100$, the form \eqref{eq:BRFriedm} with $w_m=0$ suffices (i.e., pressureless dust like baryons and cold dark matter). Occasionally, we are interested in the $z \to \infty$ limit, whereupon we set $w_m = 1/3$ (i.e., radiation). Here, we introduced the redshift $z \equiv a_0/a-1$. $\Omega_\mathrm{DE}$ is a dynamical ``dark energy" contribution due to the massive spin-2 field,
\begin{equation}
\label{eq:OmegaDefs}
	\omegade = \frac{B_0}{3} + B_1 y + B_2 y^2 + \frac{B_3}{3} y^3, \quad y \equiv \frac{\wt{a}}{ca}.
\end{equation}
See Fig.~\ref{fig:omegade} for some examples. The ratio of the scale factors, $y$, is determined by the quartic polynomial,
\begin{equation}
\label{eq:yPoly}
	- \frac{B_1}{3 \tan^2 \theta} + \left(\frac{B_0}{3} - \frac{B_2}{\tan^2 \theta} + \Omega_m\right) y + \left(B_1 - \frac{B_3}{\tan^2 \theta}\right) y^2 + \left(B_2 - \frac{B_4}{3 \tan^2 \theta}\right) y^3 + \frac{B_3}{3} y^4 = 0.
\end{equation}
An alternative to equation \eqref{eq:BRFriedm} can be found by eliminating $\Omega_m$, using \eqref{eq:yPoly},
\begin{equation}
\label{eq:BRFriedm_Alt}
E^2 = \omegadef + \Omega_k, \quad \omegadef \equiv \frac{1}{\tan^2 \theta \, y^2} \left(\frac{B_1}{3} y + B_2 y^2 + B_3 y^3 + \frac{B_4}{3} y^4\right).
\end{equation}
Thus, in a spatially flat universe $\Omega_k=0$, the expansion rate can be expressed analytically in terms of the ratio of the scale factors, $E=E(y)$. Here, $\omegadef$ is the dark energy in the $f$-sector, up to a factor of $(y_0 / y)^2$, which can be shown by noting that the Hubble parameter of $\f$ can be written $\wt{H}=H/y$. To close the system of equations, besides \eqref{eq:BRFriedm} and \eqref{eq:yPoly}, we have conservation of matter stress--energy and the matter equation of state, assuming a perfect fluid with constant equation of state parameter $w_m$,
\begin{equation}
\label{eq:OmegaMcons}
\Omega'_m \equiv \frac{d \Omega_m}{d \ln a}= -3 (1+w_m) \Omega_m, \quad \rho = w_m P.
\end{equation}
Prime denotes derivative with respect to $e$-folds, $'=d/d \ln a$. The equation of state parameter for the dark energy $w_\mathrm{DE}$ is time dependent and defined via,
\begin{equation}
	\omegade' = -3 (1+ w_\mathrm{DE}) \omegade.
\end{equation}
From this, one can show that,
\begin{equation}
\label{eq:wDE_Alt}
	w_\mathrm{DE} = -1 - \frac{1}{3} y' \frac{d \ln \omegade}{dy}.
\end{equation}
The Universe today is expanding, even accelerating, which is shown by supernova type Ia measurements \cite{Riess:1998cb,Perlmutter:1998np}, hence $H_0 >0$. Being conservative, we will assume that the Universe has been expanding throughout its history, that is $H>0$ and hence $\Omega_m$ and $z$ are decreasing with time (assuming $w_m > -1$) while $a$ is increasing. As time goes, $\Omega_m$ decreases and in the limit $\Omega_m \to 0$, $y$ is constant, as can be seen from \eqref{eq:yPoly}. In fact, it is equal to one, $y|_{t \to \infty} = 1$ which follows from its definition \eqref{eq:OmegaDefs}. Hence, in the asymptotic future, bimetric cosmology tends to a proportional dS solution with a cosmological constant,
\begin{equation}
\label{eq:Omegag}
\omegaleff \equiv \omegade|_{t \to \infty} = \frac{B_0}{3} + B_1 + B_2 + \frac{B_3}{3},
\end{equation}
completing our set of five physical parameters. Note that the physical parameter $\omegaleff$ is independent of the rescaling invariance.

In the final de Sitter phase, one must require,
\begin{equation}
\label{eq:Higuchi1}
\mfphat^2 > 2 \omegaleff,
\end{equation}
in order to avoid the Higuchi ghost \cite{Higuchi:1986py}. The constraint can be generalized to dynamical cosmological backgrounds \cite{Fasiello:2013woa}, including spatial curvature \cite{DeFelice:2014nja},
\begin{equation}
\label{eq:dynHiguchi}
\meffhat^2 > 2 \omegadef,
\end{equation}
where we have introduced a dynamical graviton mass $\meffhat(z)$,
\begin{equation}
\label{eq:mEff}
\meffhat^2 \equiv \left(1 + \frac{1}{\tan^2 \theta \, y^2}\right) (B_1 y + 2 B_2 y^2 + B_3 y^3).
\end{equation}
$m_\mathrm{eff}$ coincides with the Fierz--Pauli mass $\mfphat$ in the final dS phase where $\omegadef = \omegade$ and $y=1$.

Two useful identities are,
\begin{align}
	\label{eq:dOmegaDE_dy}
	y \frac{d \omegade}{dy} &= \left(1 + \frac{1}{\tan^2 \theta \, y^2}\right)^{-1} \meffhat^2,\\
	\label{eq:dOmegam_dy}
	-y \frac{d\Omega_m}{dy} &= \meffhat^2 -2 \omegadef.
\end{align}
The equations can be checked by inserting the definitions of $\omegade$, $\omegadef$ and $\meffhat$. We will use these relations to show that $y$ and $\omegade$ are monotonically increasing with redshift. 

Since the polynomial for $y$ \eqref{eq:yPoly} is quartic, it has a closed-form solution with up to four real solutions. However, only one of them is consistent, as discussed below.

\subsection{Infinite branch (inconsistent)}
These solutions to \eqref{eq:yPoly} have an infinite range in $y$ with $y$ diverging with increasing redshift (i.e., backwards in time, or increasing $\Omega_m$) \cite{Konnig:2013gxa}. These solutions are plagued by a Higuchi ghost and are therefore ruled out, see Section~\ref{sec:TheorConstr}. Moreover, in the early time limit, the quartic polynomial for $y$ \eqref{eq:yPoly} is to leading order (assuming $B_3 \neq 0$),
\begin{equation}
	\Omega_m y + \frac{B_3}{3} y^4 \simeq  0, \quad y,\Omega_m \to \infty,
\end{equation}
and thus for $y$ and $\Omega_m$ to be positive, we need $B_3 \leq 0$. This excludes a working Vainshtein mechanism which requires $B_3 > 0$, see Section~\ref{sec:TheorConstr}.

\subsection{Finite branch (consistent)}
These solutions to \eqref{eq:yPoly} have a finite range in $y$ with $y \to 0$ as $\Omega_m \to \infty$ (i.e., as $z \to \infty$). In this early-time limit, the massive spin-2 field contributes with an effective cosmological constant energy density $\omegade|_{z \to \infty} = B_0 / 3$, hence the dark energy equation of state $w_\mathrm{DE} \to -1$ as $z \to \infty$, which can be seen from \eqref{eq:BRFriedm} and \eqref{eq:OmegaDefs}. The exceptions are the self-accelerating models (i.e., $B_0=0$), for which $w_\mathrm{DE} \to -(2+w_m)$ as $z \to \infty$. According to \eqref{eq:BetaToPhys}, the early universe cosmological constant is negative for large $\mfphat$, $\alpha$, or $\beta$. See Fig.~\ref{fig:omegade} (left panel) for some examples.

The difference between the cosmological constants of the early universe and the late universe is,
\begin{equation}
\label{eq:DeltaOmegaL}
\Delta \Omega_\Lambda \equiv \omegade|_{z = -1} - \omegade|_{z \to \infty} = \frac{1}{3} \sin^2 \theta \, \mfphat^2 (3 + 3\alpha + \beta).
\end{equation}
See Fig.~\ref{fig:omegade} for examples in the case of general models and self-accelerating models. For large (positive) values of $\mfphat$, $\alpha$, and $\beta$, we have a negative cosmological constant in the early universe, $\omegade|_{z \to \infty} <0$. Large negative $\alpha$ or $\beta$ have the opposite effect. (Note however that such values are precluded by a working Vainshtein mechanism, see Section~\ref{sec:TheorConstr}.)

\begin{figure}[t]
	\centering
	\includegraphics[width=\linewidth]{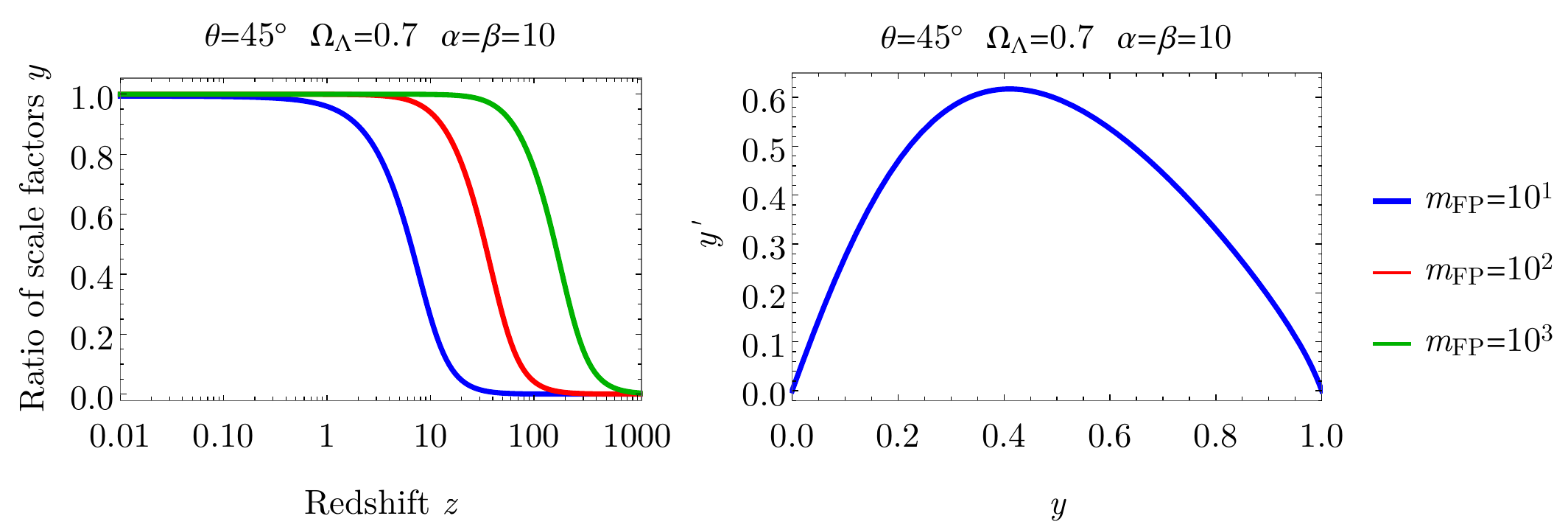}
	\caption{\emph{Left panel}: Ratio of the scale factors $y = \wt{a}/(ca)$ as a function of redshift. The ratio starts at $y=0$ at the Big Bang, increases monotonically with time, and ends up at $y=1$ in the future infinity. Increasing $\mfphat$ pushes the transition between the late-time and early-time plateaus towards higher redshifts. The same holds for increasing values of $\alpha$ and $\beta$. \emph{Right panel}: The phase space of $y$. Since the curves with $\mfphat = 10^1,10^2,10^3$ have the same shape (see the left panel), their phase space curves are virtually overlapping, which is why only $\mfphat=10^1$ is visible in the right panel.}
	\label{fig:yexamples}
\end{figure}

From the dynamical Higuchi bound \eqref{eq:dynHiguchi} and \eqref{eq:dOmegam_dy}, it follows that $y$ increases monotonically with time ($y'>0$) from $y=0$ at the Big Bang to $y=1$ in the asymptotic future, see Fig.~\ref{fig:yexamples}. In other words, $y$ can be used as a time coordinate. Observations support $\Omega_k \simeq 0$, not only for a $\Lambda$CDM model but also for bimetric models \cite{Lindner:2020eez}. Setting $\Omega_k = 0$ henceforth, the dynamical Higuchi bound gives $\meffhat^2 > 2 E^2 >0$ and from \eqref{eq:dOmegaDE_dy}, it then follows that $\omegade$ increases monotonically with time, see Fig.~\ref{fig:omegade}. 

\noindent From \eqref{eq:wDE_Alt}, it follows that the equation of state is $w_\mathrm{DE}<-1$ if $\omegade>0$ (i.e., a phantom equation of state) and $w_\mathrm{DE}>-1$ if $\omegade < 0$. This implies that the (effective) dark energy fluid always violates the null, dominant, weak and strong energy conditions. The violation of the dominant energy condition means that the bimetric fluid propagates superluminally with respect to the physical metric, but in a way which does not violate causality \cite{Hassan:2017ugh}. The phantom equation of state does not lead to a Big Rip within the finite future since $w_\mathrm{DE}$ is time-dependent and tends to a cosmological constant $w_\mathrm{DE} \to -1$ fast enough in the late universe, which we show analytically in Appendix~\ref{sec:BigRip}, see also Fig.~\ref{fig:wde}. 

\begin{figure}[t]
	\centering
	\includegraphics[width=0.9\linewidth]{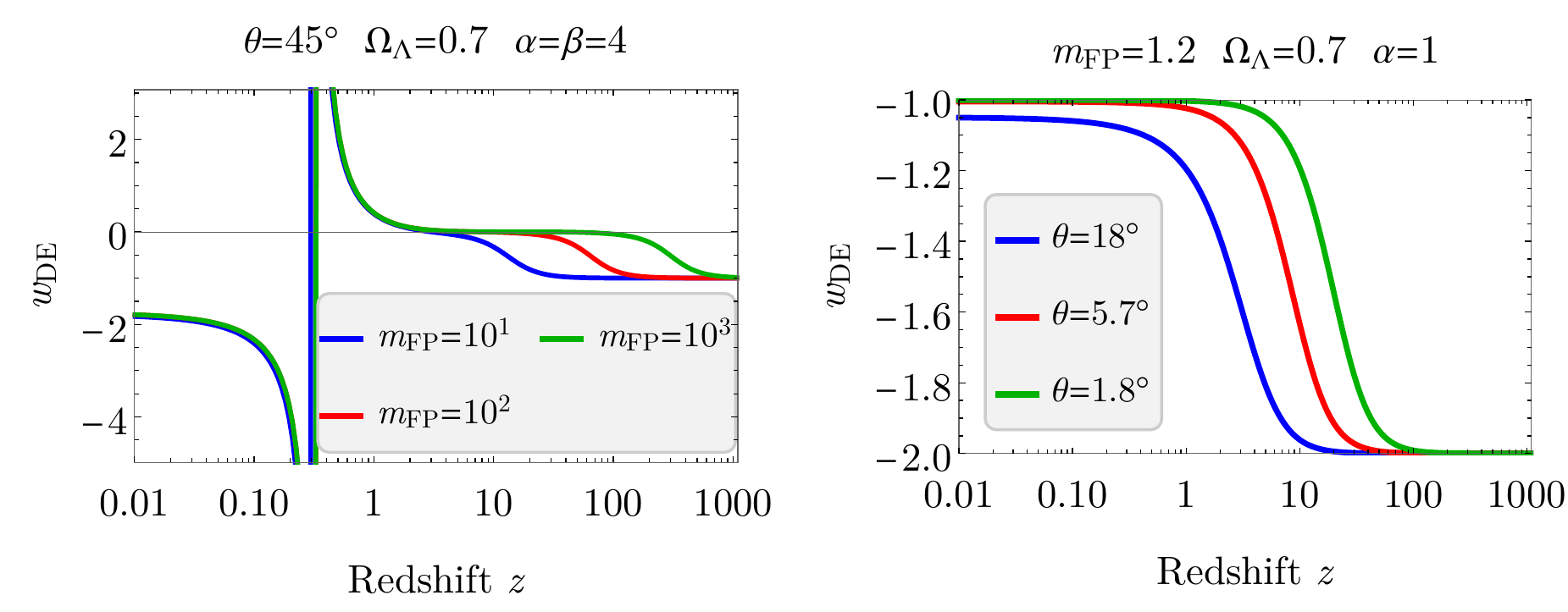}
	\caption{Equation of state (EoS) for the bimetric dark energy fluid, as a function of redshift. Here, we have set $w_m =0$. \emph{Left panel}: general models with increasing $\mfphat$. \emph{Right panel}: self-accelerating models with decreasing $\theta$. For the general models, the EoS diverges when (and if) the dark energy density passes through zero, as can be seen from \eqref{eq:wDE_Alt}. There is no physical singularity at this point and the cosmology is well-behaved. For a self-accelerating model, the dark energy fluid can be divided into two phases: (i) early-time with phantom EoS $w_\mathrm{DE}=-(2+w_m)$ and (ii) late-time cosmological constant phase $w_\mathrm{DE}=-1$. For a general model, we identify the following phases: (1) early-time cosmological constant phase $w_\mathrm{DE}=-1$, and (2) late-time cosmological constant phase $w_\mathrm{DE}=-1$. In the examples in the left panel, we can also identify an intermediate region where the dark energy fluid acts as dark matter. Increasing $\mfphat$, this phase becomes extended towards higher redshifts. This is also the case with increasing $\alpha$ or $\beta$. It should be stressed that the intermediate region between the early-time and late-time cosmological constant phases can assume many different shapes.}
	\label{fig:wde}
\end{figure}

To summarize, the finite branch cosmology is consistent and is defined by choosing the lowest lying, strictly positive root of the quartic equation \eqref{eq:yPoly} \cite{Luben:2020xll}. The early Universe is described by a $\Lambda$CDM model with cosmological constant energy density $\omegade|_{z \to \infty} = B_0 / 3$. For self-accelerating models $B_0=0$ and the cosmological constant vanishes in the early universe. The late universe is described by a $\Lambda$CDM model with cosmological constant $\omegaleff$ which is greater than the early-universe one. The transition between these two GR regions is due to the massive spin-2 field. There are convenient ways to solve for the cosmological evolution graphically, see \cite{Volkov:2011an,Mortsell:2016too,Mortsell:2017fog,Hogas:2019ywm}.

\newpage
\section{General relativity limits}
\label{sec:GRlims}
At the level of the action \eqref{eq:HRaction}, it is evident that the theory reduces to two decoupled copies of GR if $\beta_1=\beta_2=\beta_3=0$ (i.e., $B_1 = B_2 = B_3 =0$). In terms of the physical parameters \eqref{eq:BetaToPhys}, this is the case if and only if $\theta = 0$ or $\mfphat = 0$. Taking the limit $\mfphat \to 0$ is problematic since the Higuchi ghost is excited for small values of the graviton mass (assuming that the effective cosmological constant is non-vanishing). Hence, we expect $\theta \to 0$ to be the only GR limit for the full nonlinear theory. Nevertheless, if we look at a particular type of solution, there can be other GR limits for that particular solution. It turns out that for both the local and the cosmological solutions, besides $\theta \to 0$, we approach general relativity for large values of $\mfphat$, $\alpha$, or $\beta$. For the cosmological solutions, this is true assuming that we do not impose any independent value on the matter density today $\omegamnot$.

\subsection{Local solutions}
\label{sec:GRlimLocal}
First, note that in the limit $\mfphat \to 0$, the Vainshtein radius $r_V$ increases indefinitely \eqref{eq:VainshteinDef} and the local solutions reduce to GR. However, appearance of the Higuchi ghost invalidates this limit.

Second, in the large graviton mass limit $\mfphat \to \infty$, the linearized solution is applicable and since the Compton wavelength is much smaller than the size of the system, general relativity results are restored \cite{Luben:2018ekw}. This is true for a general bimetric model. However, for self-accelerating models, a large value of the graviton mass violates the cosmological constraints of Section~\ref{sec:TheorConstr} (see also Fig.~\ref{fig:dynhig2}), unless we also set a small value of the mixing angle $\theta$. As an example, if $\mfp \gtrsim 10^{33}$ the Yukawa term of the linearized solutions is suppressed enough to retain GR results within the experimental precision of solar system tests \cite{Luben:2018ekw}. However, the cosmological constraints of Section~\ref{sec:TheorConstr} then implies $\theta \lesssim 10^{-34}$.

\begin{figure}[t]
	\centering
	\includegraphics[width=\linewidth]{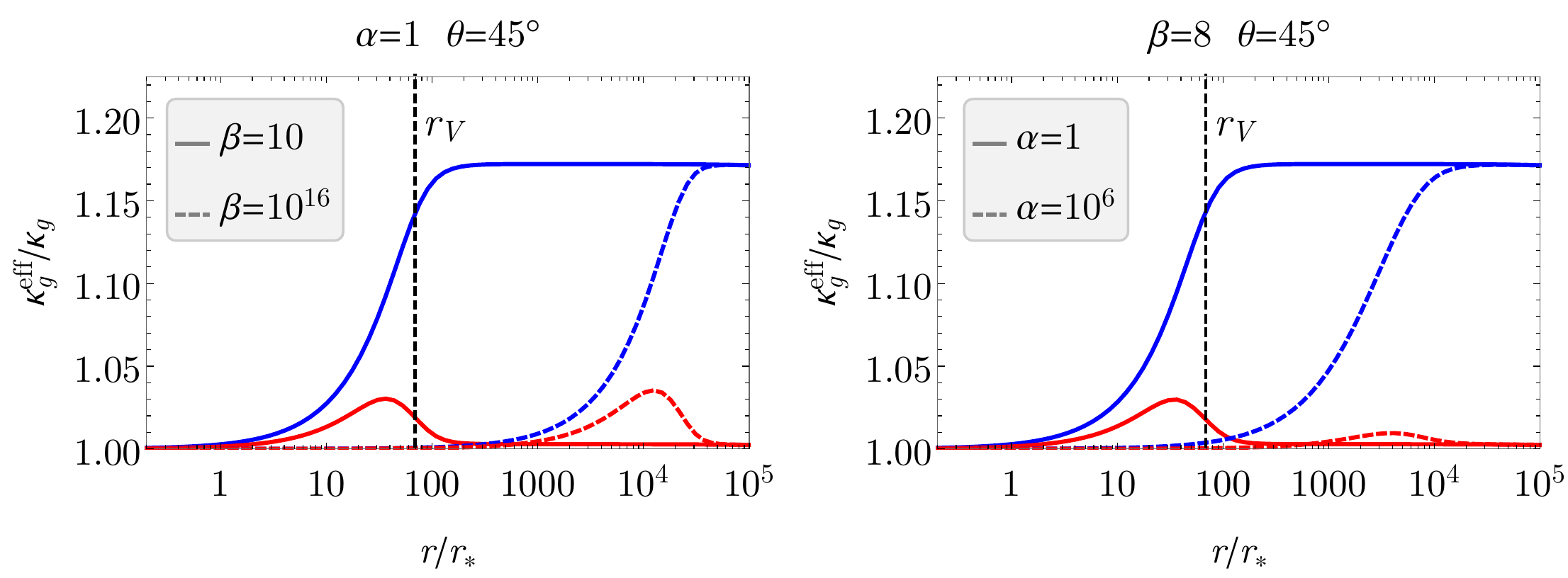}
	\caption{Effective gravitational ``constants" for massive particles (blue) and massless particles (red). The gravitational constant for massive particles is constant and equal to $\kappa_g$ for small radii and constant and equal to $\kappa_g(1+\sin^2 \theta /3)$ for large radii (assuming that we are well inside the Compton wavelength $\lambda_g$ of the graviton). Between these regions, there is a transition. The gravitational constant for massless particles is equal to $\kappa_g$ both for small radii and large radii (assuming $r \ll \lambda_g$). In between these regions, there is a peak. \emph{Left panel}: Increasing values of $\beta$. For solid curves, $\beta=10$, and for dashed curves, $\beta=10^{16}$. \emph{Right panel}: increasing values of $\alpha$. For solid curves, $\alpha=1$, and for dashed curves $\alpha=10^6$. The transition region moves outwards with increasing $\alpha$ or $\beta$. In other words, for large $\alpha$ or $\beta$, GR is restored also well outside $r_V$.}
	\label{fig:incralphabeta}
\end{figure}

If we expand the polynomial for $\mu$ \eqref{eq:muPoly} around $1 / \alpha =0$ (i.e., in the $\alpha \to \infty$ limit), we see that $\mu \to 0$ and $\alpha^2 \mu^3 \to 0$. From \eqref{eq:VainsteinPots}, it follows that the transition to the $r\gg r_V$ plateau is shifted outwards indefinitely with increasing $\alpha$ but the final value of the potentials on the plateau remains the same, see Fig.~\ref{fig:incralphabeta}. This means that $r_V$ as defined in \eqref{eq:VainshteinDef} sets a lower limit for $r$ inside which we start to approach GR.

Similarly, expanding \eqref{eq:muPoly} around $1/\beta = 0$, it follows that $\mu \to 0$ and $\beta \mu^3 \to 0$. Increasing $\beta$, the transition to the $r \gg r_V$ plateau is pushed outwards indefinitely, see Fig.~\ref{fig:incralphabeta}.

\subsection{Background cosmology}
If the bimetric cosmological models reduce to $\Lambda$CDM for a particular parameter choice, then we have a cosmological GR limit at the background level. Then, in particular, the bimetric models mimic $\Lambda$CDM close to the final de Sitter phase. Therefore, we can obtain a complete list of candidate GR limits by sorting out those models that reduces to $\Lambda$CDM close to $z=-1$. To analyze the cosmology close to this point, we expand the quartic polynomial \eqref{eq:yPoly} around small $\Omega_m$ and solve order by order in this small parameter. The expression can then be inserted in the modified Friedmann equation \eqref{eq:BRFriedm} with the result,
\begin{equation}
\label{eq:Eexp}
E^2 = \omegaleff + \frac{\kgeff}{\kappa_g} \Omega_m + (1 + 2 \alpha) \sin^2 \theta \frac{\mfphat^2 \omegaleff}{(\mfphat^2 - 2 \omegaleff)^3} \Omega_m^2 + \mathcal{O}(\Omega_m^3), \quad \Omega_m \ll 1,
\end{equation}
where,
\begin{equation}
\label{eq:kgeff}
\kgeff = \kappa_g \left(1 - \sin^2 \theta \frac{\mfphat^2 }{\mfphat^2  - 2 \omegaleff}\right),
\end{equation}
see for example \cite{Babichev:2016bxi}. For an expansion up to order $\Omega_m^4$, see Appendix~\ref{sec:dSexp}. A necessary condition to retain $\Lambda$CDM around dS (assuming that the expansion is valid) is that all $\Omega_m^n$ terms in the expansion with $n \geq 2$ vanish to any order that we choose. A word of caution: in the limits $\alpha \to \infty$ and $\beta \to \infty$, the expansion \eqref{eq:Eexp} is invalid since the higher order terms diverge. These limits have to be studied by other means, see below. From \eqref{eq:Eexp}, we can immediately identify five ways to cancel the $\Omega_m^2$ term and hence identify seven cosmological GR limit candidates.

\begin{enumerate}[ {(}1{)}]
	\item As is well known, in the limit $\theta  \to 0$, we retain GR and hence $\Lambda$CDM over the whole history of the Universe.
	
	\item Setting $\omegaleff = 0$. However, this leads to a CDM model, that is without a cosmological constant, which is observationally ruled out.
	
	\item In the limit $\mfphat \to 0$, the $\Omega_m^2$ term cancels. However, this limit is problematic since it excites the Higuchi ghost. If we set $\mfphat=0$ exactly, we recover two decoupled copies of GR (cf. \eqref{eq:HRaction}). 
	
	\item Another possibility is $\alpha = -1/2$. Proceeding to order $\Omega_m^3$, we must set $\beta = 0$. This means $B_1 = B_3 = 0$. As shown in Ref.~\cite{vonStrauss:2011mq}, this reduces to a $\Lambda$CDM model at all redshifts, with a rescaled matter density and cosmological constant. However, this model is problematic due to a singularity in the second metric at a finite redshift. Since the two metrics live on the same space-time, also the physical metric is singular at this point in time, even if its curvature scalars are finite \cite{Kocic:2017wwf,Torsello:2017ouh}. Nevertheless, very close to the $(\alpha = -1/2,\beta=0)$ point, there are models closely resembling $\Lambda$CDM which have a nonsingular evolution all the way to $z=\infty$. For example approaching $\alpha = -1/2$ from the right along the $\beta=0$ line. Note however that the neighborhood around $(\alpha = -1/2,\beta=0)$ is excluded by requiring a working Vainshtein screening mechanism, as we will show in Section~\ref{sec:TheorConstr}, see Fig.~\ref{fig:dynhig2}. 
	
	\item Infinite graviton mass, $\mfphat \to \infty$ (i.e., vanishing Compton wavelength). We checked that all terms up to (and including) $\Omega_m^5$ cancel, see also Ref. \cite{DeFelice:2013nba}.\footnote{For the expansion up to order $\Omega_m^4$, see Appendix~\ref{sec:dSexp}. We also computed the $\Omega_m^5$ term which we do not display due to its length.} This is in line with our expectation that the massive spin-2 field should manifest its dynamics when its mass is of the same order as the energy scale set by $H$.\footnote{Or, equivalently, when the Compton wavelength of the graviton is equal to the Hubble radius, that is when $\lambda_g \sim 1/H(z)$.}  
	
	\item $\alpha \to \infty$. This must be included as a candidate since the expansion \eqref{eq:Eexp} is invalid in this limit.
	
	\item $\beta \to \infty$. Must be included for the same reason as $\alpha \to \infty$.
\end{enumerate}
What remains to investigate is whether $\mfphat,\alpha,\beta \to \infty$ are $\Lambda$CDM limits for all redshifts. For large $\mfphat$, one can expand in a Taylor series around $1/\mfphat =0$ and solve the equations order by order. In the case where $\alpha$ is large, we expand in a Puiseux series $\sum_{n=0}^\infty c_n (1/\alpha)^{n/2}$ and with $\beta$ large in a Puiseux series $\sum_{n=0}^\infty c_n (1/\beta)^{n/3}$. The details can be found in Appendix~\ref{sec:dSexp}. The first terms are the same in each of the three limits,
\begin{subequations}
	\label{eq:ExpmAlphaBeta}
	\begin{align}
	E^2 &= \omegaleff + \cos^2 \theta \, \Omega_m(z) + \mathrm{higher \; orders},\\
	\label{eq:Omegam0lim}
	\omegamnot &= (1-\omegaleff)/\cos^2 \theta + \mathrm{higher \; orders},\\
	\omegade &= \omegaleff - \sin^2 \theta \, \Omega_m + \mathrm{higher \; orders}.
	\end{align}
\end{subequations}
Interestingly, the massive spin-2 field contributes with an effective cosmological constant plus a cold dark matter component with negative energy density (assuming $w_m =0$), compare with Fig.~\ref{fig:wde}. With $\omegaleff = 0.7$, the expansion rate $E(z)$ coincides with that of a $\Lambda$CDM model with the same cosmological constant although the matter density $\Omega_{m,0}$ is greater than $0.3$ and hence the matter density $\omegamnot$ is rescaled. Thus, in these limits, $\theta$ and $\omegamnot$ are degenerate at the background level and there are no constraints on them. This is referred to as the ``dark degeneracy" \cite{Kunz:2007rk}. Hence, $\mfphat,\alpha,\beta \to \infty$ are $\Lambda$CDM limits on the background level, as long as we do not impose any background independent constraint on $\omegamnot$. In each of the limits $\mfphat,\alpha,\beta \to \infty$, the dynamical dark energy phase (contained in the higher order terms) is pushed backwards in time and bimetric cosmology approaches a $\Lambda$CDM model with cosmological constant $\omegaleff$.

\begin{figure}[t]
	\centering
	\includegraphics[width=\linewidth]{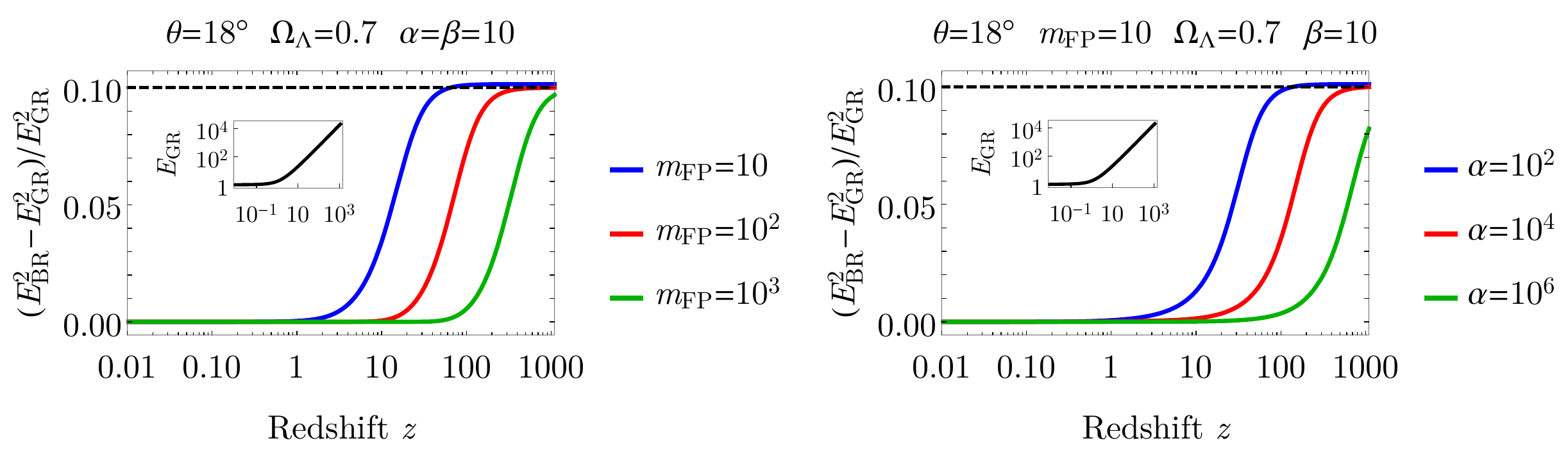}
	\caption{The relative difference in the expansion rate $E=H/H_0$ between bimetric cosmologies and a $\Lambda$CDM model, as a function of redshift. The cosmological constants of the two models are the same $\omegaleff = 0.7$. In the late universe, the bimetric models expand according to a $\Lambda$CDM model with cosmological constant $\omegaleff$ while in the early universe, it expands according to a $\Lambda$CDM model with cosmological constant $B_0/3 < \omegaleff$. \emph{Left panel}: Increasing values of the graviton mass. \emph{Right panel}: Increasing values of $\alpha$. A similar plot is obtained for increasing values of $\beta$. Increasing values of $\mfphat$, $\alpha$, or $\beta$ pushes the transition from the late-time to the early-time $\Lambda$CDM phases to higher redshifts. The early-time difference between the expansion rates of the bimetric models and $\Lambda$CDM is set by $\tan^2 \theta = 0.1$, in accordance with \eqref{eq:DeltaE}.}
	\label{fig:ediff}
\end{figure}

To study the difference between the expansion rate of a general bimetric model and a $\Lambda$CDM model, we consider the difference in the expansion rates (squared),
\begin{equation}
\Delta E^2 \equiv \Ebr^2 - \Egr^2.
\end{equation}
Since $\Ebr|_{z=0} = \Egr|_{z=0} = 1$, $\Delta E^2|_{z=0}=0$. At early times, matter dominates over dark energy and,
\begin{equation}
\label{eq:DeltaE}
\left. \frac{\Delta E^2}{E_\mathrm{GR}^2} \right|_{z\to\infty} = \frac{\Omega_m-\Omega_m^\mathrm{GR}}{\Omega_m^\mathrm{GR}} = \frac{\Omega_{m,0}}{\Omega_{m,0}^\mathrm{GR}}-1.
\end{equation}
Therefore, in the large $\mfphat$, $\alpha$, and $\beta$ limits, setting the same cosmological constants in a $\Lambda$CDM model, we get,
\begin{equation}
\label{eq:ErelDiff}
\left. \frac{\Delta E^2}{E_\mathrm{GR}^2} \right|_{z \to \infty} = \frac{(1-\Omega_\Lambda)/\cos^2 \theta }{1-\Omega_\Lambda}-1 = \tan^2 \theta >0,
\end{equation}
where we used \eqref{eq:Omegam0lim} and \eqref{eq:DeltaE}. Hence, in these limits $E > E_\mathrm{GR}$ in the early universe with the relative difference set by $\tan^2 \theta$. Remember that the expansion rate $E$ agrees with that of a $\Lambda$CDM model until a redshift $z_t$. As we take the large $\mfphat$, $\alpha$, or $\beta$ limit, this redshift is pushed backwards in time indefinitely, see Fig.~\ref{fig:ediff} for examples.

There are striking similarities between the local solutions with a working Vainshtein mechanism and the cosmological background solutions \cite{Luben:2019yyx}. In both cases, there are two GR regions: late/early universe for the cosmological solutions and close/far away from the source for the local solutions, and the length scale of the transition between the two is set by the Compton wavelength of the massive graviton (i.e., the graviton mass). For both the local and cosmological solutions, $\alpha \to \infty$ and $\beta \to \infty$ are GR limits in the sense of extending the transition to greater length scales or earlier times, compare Fig.~\ref{fig:incralphabeta} with Fig.~\ref{fig:ediff}. In the limit $\mfphat \to \infty$, $\Lambda$CDM cosmology is recovered. For the local solutions, in this limit the linearized solutions \eqref{eq:LinPot} are valid for a general bimetric model (see Section~\ref{sec:GRlimLocal} for a caveat concerning the self-accelerating models) and the exponential Yukawa terms become suppressed, restoring GR locally.

\section{Constraints on the physical parameters}
\label{sec:TheorConstr}
The constraints due to the local solutions and cosmological background, that we will present here, are shown together in Fig.~\ref{fig:dynhig2}.

To have the correct sign on the kinetic term for $\f$ \eqref{eq:HRaction}, we require a real-valued mixing angle $\theta$, corresponding to $\kappa_g / \kappa_f > 0$, see \eqref{eq:kbarDef}. Linearly stable proportional solutions demands a real-valued graviton mass, $\mfphat^2 > 0$ \cite{Hassan:2012wr}.

\subsection{Local solutions}
\begin{figure}[t]
	\centering
	\includegraphics[width=0.55\linewidth]{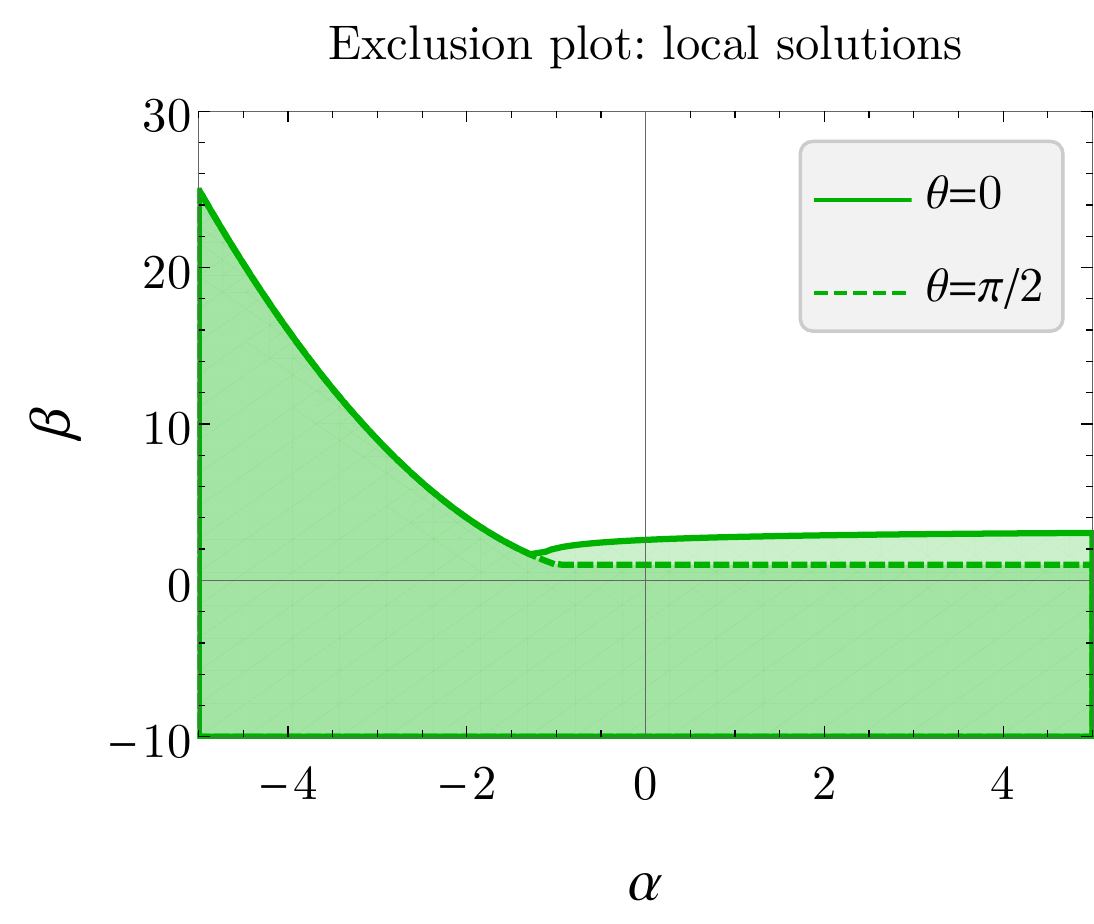}
	\caption{Exclusion plot in the $\alpha \beta$-plane from demanding a working Vainshtein mechanism. We plot the extreme cases $\theta=0$, which gives the most restrictive constraints, and $\theta = \pi / 2$, which gives the least restrictive constraints. For $\alpha \lesssim -3/2$, the constraints coincide.}
	\label{fig:vainshtein}
\end{figure}

To have a working Vainshtein mechanism, it is necessary and sufficient to require \cite{Enander:2015kda},\footnote{Note that there are two typos in Ref.~\cite{Enander:2015kda}. First, in the sentence before their eq. (4.15), the constraint should read $\alpha > - \sqrt{\beta}$. This correction holds everywhere else where this constraint is discussed. Also, in their eq. (4.15), the definitions for $d_1$ and $d_2$ should be interchanged.}
\begin{subequations}
	\label{eq:VainshteinConstraints}
	\begin{align}
	\label{eq:VainshteinBeta}
	\beta &> 1,\\
	\label{eq:VainshteinAlpha}
	\alpha &> - \sqrt{\beta},\\
	\label{eq:VainshteinGeneral}
	(\alpha + \sqrt{\beta}) \left(\alpha d_1 + d_2\right)&>0,
	\end{align}
\end{subequations}
where,
\begin{subequations}
	\begin{align}
	d_1 &\equiv 1 + 3 \tan^2 \theta -6 \sqrt{\beta} \sec^2 \theta + 3 \beta \sec^2 \theta,\\
	d_2 &\equiv -1 + 6\sqrt{\beta}(1+\beta) \sec^2 \theta - \beta (13 + 12 \tan^2 \theta),
	\end{align}
\end{subequations}
see Fig.~\ref{fig:vainshtein}. These constraints rely on an analytical study of the solutions to the seventh degree polynomial \eqref{eq:muPoly}. We also implemented a numerical algorithm, scanning the parameter space for points with working Vainshtein mechanism. For each point in the $\alpha\beta$-plane, the algorithm solves numerically for $\mu(r)$ (in fact, it solves for $r(\mu)$) and accepts the solution if it is real valued and increasing with $r$. The analytical and numerical constraints agree.

Note that \eqref{eq:VainshteinBeta} implies $B_3 >0$ and that $B_1>0$ (obtained from the background cosmology, see \eqref{eq:PosBeta1}) together with \eqref{eq:VainshteinBeta} implies $B_2<0$, that is,
\begin{equation}
	B_1 >0, \quad B_2 <0, \quad B_3 >0.
\end{equation}
In other words, to have a viable background cosmology and working Vainshtein screening mechanism, we must include all the bimetric interaction parameters (except the cosmological constants $B_0$ and $B_4$). Also note that $B_3>0$ excludes the infinite branch cosmology which becomes imaginary at early times in this parameter range.

\subsection{Background cosmology}
\begin{figure}[t]
	\centering
	\includegraphics[width=\linewidth]{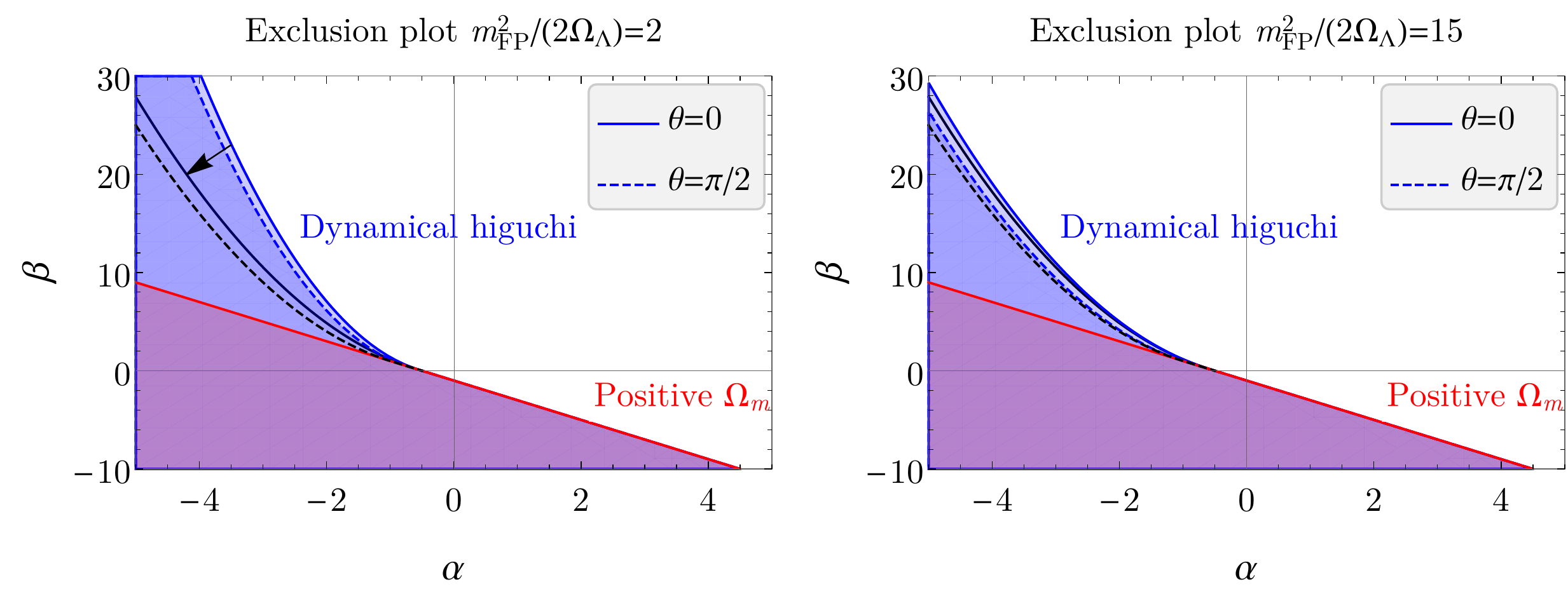}
	\caption{Exclusion plot in the $\alpha\beta$-plane from background cosmology constraints. The red/purple region is excluded according to the requirement that $\Omega_m$ is positive in the early universe. The blue region is excluded due to the dynamical Higuchi bound. The solid blue boundary is for $\theta=0$ and the dashed blue boundary is for $\theta = \pi / 2$. The arrow indicates the direction in which the boundary moves with increasing $\mfphat^2/(2\omegaleff)$. In the left panel $\mfphat^2/(2\omegaleff)=2$ and in the right panel $\mfphat^2/(2\omegaleff)=15$. The black curves denote the boundaries for $\theta=0$ (solid) and $\theta = \pi /2$ (dashed) in the limit $\mfphat^2/(2\omegaleff) \to \infty$.}
	\label{fig:dynhig3}
\end{figure}

In order for $\Omega_m$ to be positive in the early universe (i.e., as $y \to 0$), $B_1 >0$, see \eqref{eq:yPoly}. In terms of the physical parameters,
\begin{equation}
\label{eq:PosBeta1}
1 + 2 \alpha + \beta>0.
\end{equation}
To have a real-valued cosmology in the final de Sitter phase where $E^2 = \omegaleff$,
\begin{equation}
\label{eq:omegaleffPos}
\omegaleff >0.
\end{equation}
The dynamical Higuchi bound is (allowing also for non-zero spatial curvature),
\begin{equation}
	\meffhat^2 > 2\omegadef,
\end{equation}
or equivalently, using \eqref{eq:dOmegam_dy},
\begin{equation}
\label{eq:HiguchiOmegam}
	\frac{d \Omega_m}{dy}<0.
\end{equation}
With $\Omega_m(z)$ being a continuous function of redshift, the dynamical Higuchi bound guarantees that $y$ is also a continuous function of redshift and, thereby, also $\omegade(z)$ and $E(z)$. Using \eqref{eq:yPoly}, we can express $\Omega_m$ in terms of $y$ and the physical parameters and eq. \eqref{eq:HiguchiOmegam} then reads,
\begin{equation}
	\label{eq:DynHig3}
	\frac{\mfphat^2}{2 \omegaleff} > R(y;\theta,\alpha,\beta),
\end{equation}
where the ratio $R$ is,
\begin{subequations}
	\begin{alignat}{2}
		R &\equiv& &N/D,\\
		N &\equiv& &3 \sec^2 \theta \, y^3,\\
		D &\equiv& &1+2\alpha+\beta + 3 [\tan^2 \theta (1+2\alpha+\beta) - \beta] y^2 +\\ \nonumber
		& & & + 2[1-\alpha+\beta-3 \tan^2 \theta (\alpha+\beta)] y^3 + 3 \tan^2 \theta \, \beta y^4.
	\end{alignat}
\end{subequations}
The inequality \eqref{eq:DynHig3} must hold for all $y \in [0,1]$. Since $R|_{y=1}=1$, eq. \eqref{eq:DynHig3} reduces to the ordinary Highuchi constraint at the final de Sitter point,
\begin{equation}
\label{eq:Higuchi}
	\mfphat^2 > 2 \omegaleff.
\end{equation}
The dynamical Higuchi constraint can be expressed as two inequalities of the form $\beta > f_1(\theta,\alpha)$ and $\mfphat/2\omegaleff > f_2(\theta,\alpha,\beta)$, see Fig.~\ref{fig:dynhig3} for examples and Appendix~\ref{sec:Higuchi} for details. The functions $f_1$ and $f_2$ assume different forms depending on whether $\alpha >-1/2$ or $\alpha < -1/2$. As an example, in the case $\theta=0$, the constraints read,
\begin{subequations}
	\begin{alignat}{2}
		\beta|_{\theta=0} &> -1 -2\alpha,& \quad \alpha &> -1/2,\\
		\beta|_{\theta=0} &> \frac{1}{6} (\alpha-1) \left(3 + 3\alpha - \sqrt{9\alpha^2 - 6\alpha - 3}\right),& \quad \alpha &< -1/2,\\
		\mfphat^2 / (2\omegaleff)|_{\theta=0} &> 1,& \quad \alpha &> -1/2,\\
		\mfphat^2 / (2\omegaleff)|_{\theta=0} &> \frac{3}{2} \left(1 - \alpha + \beta \left[1 - \sqrt{\frac{\beta}{1+2\alpha+\beta}}\right]\right)^{-1},& \quad \alpha &< -1/2,.
	\end{alignat}
\end{subequations}

\begin{figure}[h]
	\centering
	\includegraphics[width=0.9\linewidth]{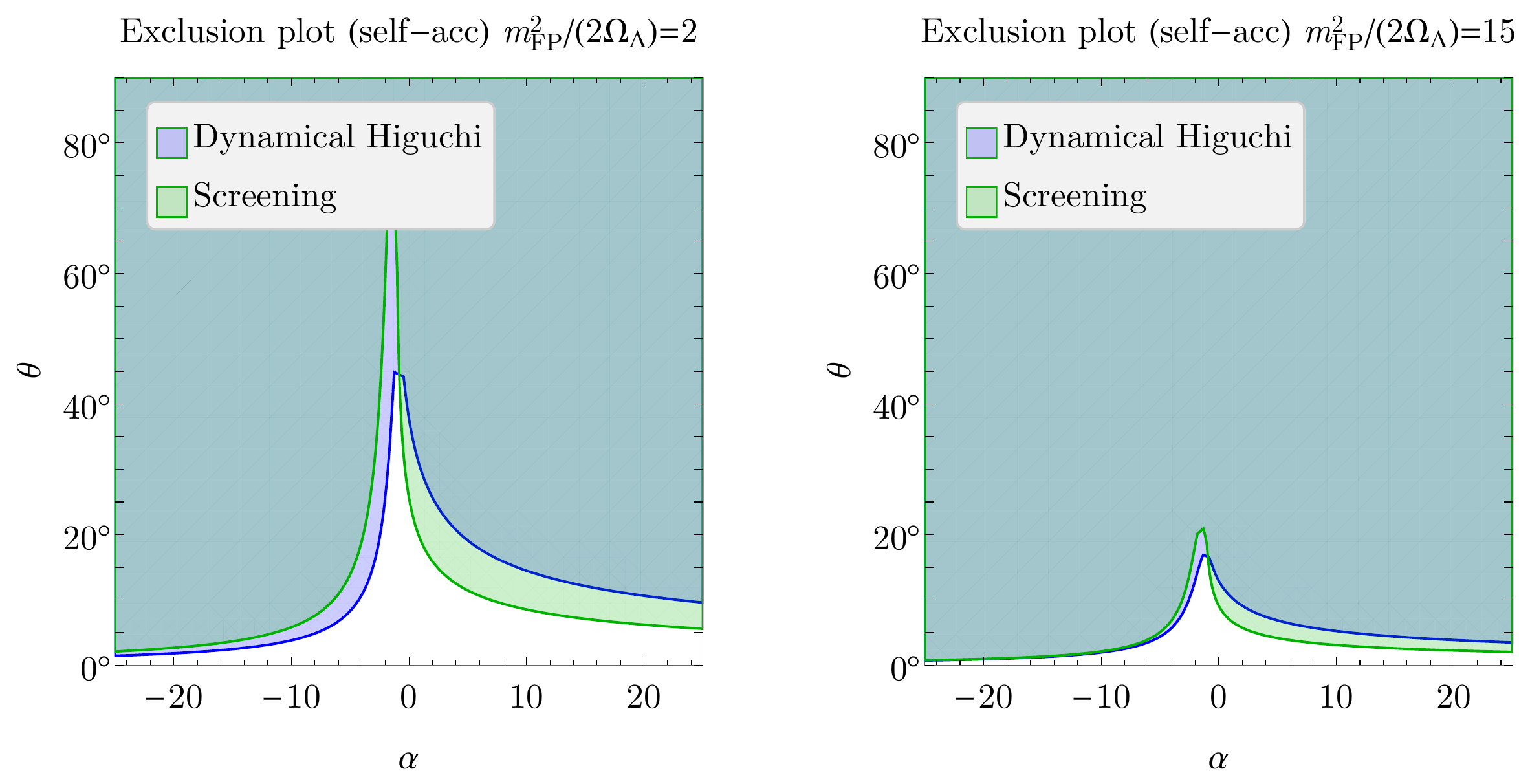}
	\caption{Exclusion plot in the $\alpha \theta$-plane for self-accelerating models. The two panels are for two different values of $\mfphat^2 /(2\omegaleff)$. Note the different scales on the vertical axes.}
	\label{fig:ConstrSelfAcc}
\end{figure}

\noindent In particular, for self-accelerating models, a large value of $\theta$ (i.e., close to $\pi /2$) require $\mfphat^2/(2\omegaleff)\simeq 1$. For a fixed value of $\mfphat^2/(2\omegaleff)$, there is an upper bound for $\theta$. Fixing also $\theta$ there is a finite range of allowed values for $\alpha$. See Fig.~\ref{fig:ConstrSelfAcc}.

\begin{figure}
	\centering
	\includegraphics[width=1.0\linewidth]{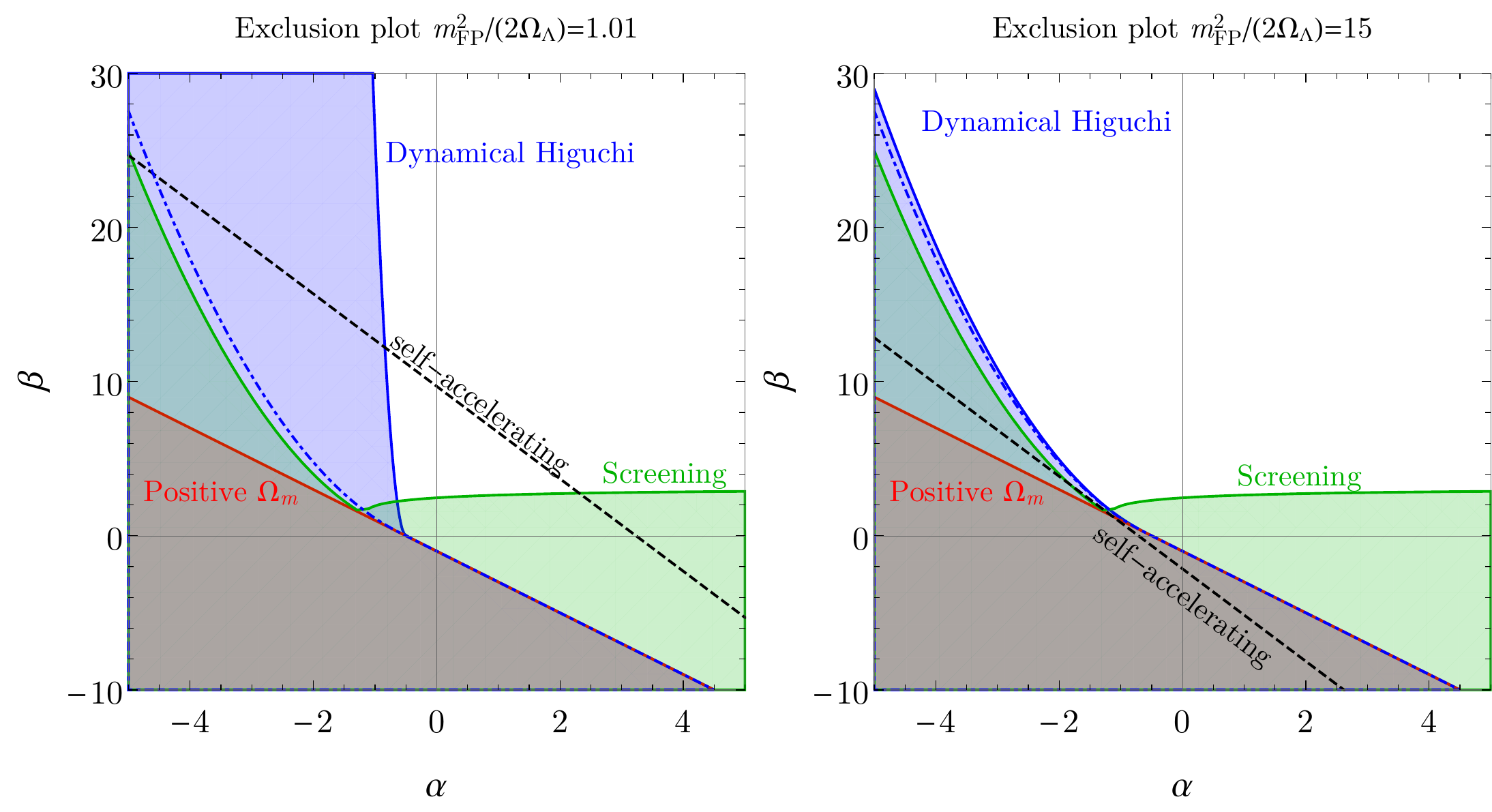}
	\caption{Exclusion plot in the $\alpha \beta$-plane from demanding a working Vainshtein screening mechanism (green region), the dynamical Higuchi bound which ensures a continuous cosmology (blue region), and positive matter density (red region). In the limit $\mfphat^2 / 2 \omegaleff \to \infty$, the boundary of the blue region approaches the dash-dotted curve. The exclusion regions change only weakly with $\theta$. Here, we have set $\theta = 20^{\circ}$. The self-accelerating cosmological models lie along the dashed lines. The slope of these are always the same ($\beta = -3 \alpha + \mathrm{const.}$) but they get shifted downwards (upwards) with increasing (decreasing) $\mfphat^2/(2\omegaleff)$ or increasing (decreasing) $\theta$. For example, with the parameter values in the right panel (i.e., with a higher value of $\mfphat^2 / 2 \omegaleff$), there are no consistent self-accelerating cosmologies. If any of the parameters $\mfphat$, $\alpha$, or $\beta$ are large, the mixing angle $\theta$ must be very small for self-accelerating solutions to exist. The region above the dashed line has negative cosmological constant in the early universe $\omegade|_{z \to \infty}<0$. The region below it has positive cosmological constant in the early universe $\omegade|_{z \to \infty}>0$.}
	\label{fig:dynhig2}
\end{figure}

Summarizing so far, the theoretical constraints from background cosmology are obtained requiring a positive matter density in the early universe and the dynamical Higuchi bound. In Fig.~\ref{fig:dynhig2}, we show the cosmological and local constraints together for general models. In Fig. \ref{fig:ConstrSelfAcc}, we plot these constraints in the case of self-accelerating models.

Assuming that we have a solution to the equations of motion at some initial point (which we set to $z=0$), the above constraints ensure that the expansion history is nonsingular (except at the Big Bang). Requiring that there is a finite branch solution at $z=0$ leads to the last constraint. Evaluating the $y$ polynomial \eqref{eq:yPoly} and the modified Friedmann equation \eqref{eq:BRFriedm_Alt} at $z=0$, we get equations for $y_0$ and $\omegamnot$,
\begin{subequations}
	\begin{align}
	- \frac{B_1}{3 \tan^2 \theta } + \left(\frac{B_0}{3} - \frac{B_2}{\tan^2 \theta} + \Omega_{m,0}\right) y_0 + \left(B_1 - \frac{B_3}{\tan^2 \theta}\right) y_0^2 + \left(B_2 - \frac{B_4}{3 \tan^2 \theta}\right) y_0^3 + \frac{B_3}{3} y_0^4 &= 0,\\
	\label{eq:NormOmegaDEf}
	\omegadef(y_0) -1 &=0.
	\end{align}
\end{subequations}
Here, the $B$-parameters are shorthand expressions for the physical parameters according to \eqref{eq:BetaToPhys}. A finite branch solution at $z=0$ requires $0 < y_0 < 1$ and $\omegamnot > 0$. Writing \eqref{eq:NormOmegaDEf} in terms of the physical parameters at $y_0 =0$ and $y_0 =1$ gives,
\begin{subequations}
	\begin{align}
		\omegadef(y_0=0) -1 &= \frac{1}{3} \mfphat^2 (1+2\alpha+\beta) \cos^2 \theta>0,\\
		\omegadef(y_0=1) -1 &= \omegaleff -1.
	\end{align}
\end{subequations}
$\omegadef(y_0=1)<0$ if $\omegaleff <1$, and since $\omegadef(y_0=0)>0$ there must exist a solution for $y_0$ in the range $0 < y_0 <1$. Also, since $\Omega_{\mathrm{DE},0} < \omegaleff < 1$ we are guaranteed that $\omegamnot >0$. So, for $\omegaleff<1$ there always exist a finite branch solution for $y_0$ with $\omegamnot>0$. If $\omegaleff > 1$, the physical parameters must be constrained. In particular, we cannot be close to any of the GR limits since a spatially flat $\Lambda$CDM model does not allow a cosmological constant greater than one. Here, we do not work out the detailed form of these constraints since $\omegaleff > 1$ is observationally excluded \cite{PhysParamObs}.

\section{Discussion and outlook}
We have investigated a set of analytical constraints which guarantee that the theory has viable properties. Our results are summarized in Figs.~\ref{fig:dynhig2} and \ref{fig:ConstrSelfAcc}. A complete list of the constraints can be found in Appendix~\ref{sec:ConstraintsList}. One should note that it may be possible to obtain the properties guaranteed by the constraints in other ways. Concerning $\mfphat^2 >0$, which is needed for stable propagation of gravitational waves on proportional backgrounds $\g = c^2 \f$, one may argue that proportional solutions are non-generic in bimetric theory and that there are other ways to obtain for example flat background solutions on which gravitational waves can propagate \cite{Kocic:2017wwf}. It would be interesting to analyze the stability conditions on those solutions. We derived constraints on $\alpha$ and $\beta$ by demanding a working Vainshtein mechanism of the type presented in the paper. However, there may be other ways to achieve a working screening mechanism. To have real-valued early-time cosmology, we required $1 + 2\alpha + \beta >0$, assuming that both metrics share the same symmetries (homogeneity and isotropy). However, there may be solutions which are homogeneous and isotropic only with respect to the physical metric with no special (or different) symmetry on the second metric. Possibly, the constraint which is most difficult to relieve is a real mixing angle $\theta$ (i.e., $\kappa_g / \kappa_f >0$) since that would imply a negative sign kinetic term in the action \eqref{eq:HRaction}.

As a next step, in Ref.~\cite{PhysParamObs} we analyze whether the theory is observationally viable and what parameter regions that are preferred. We do this by fitting to cosmological data from the cosmic microwave background, baryon acoustic oscillations, and type Ia supernovae. The cosmological best-fit parameters are compatible with solar system tests and gravitational lensing from galaxies due to a working Vainshtein screening mechanism. 

Perhaps one of the most important issues to address in the future is how to deal with structure formation/cosmological perturbations in bimetric gravity. One suggestion is that nonlinear perturbations become influential when linear perturbation theory becomes unstable, thereby restoring general relativity nonlinearly \cite{Mortsell:2015exa,Luben:2019yyx}. This would be a cosmological version of the Vainshtein mechanism. Another possibility is to solve the full nonlinear equations of motion numerically, but for that the equations of motion has be written in a suitable (well-posed) form, which is a challenging task \cite{Kocic:2018ddp,Kocic:2018yvr,Torsello:2019jdg,Torsello:2019tgc,Torsello:2020wtu}. Recently, in Ref.~\cite{DeFelice:2020ecp}, an alternative form of bimetric gravity was proposed, dubbed the minimal theory of bigravity. This theory breaks space-time diffeomorphism invariance down to the spatial subgroup and propagates four degrees of freedom. Interestingly, its background cosmology is identical to standard bimetric gravity and hence all cosmological results of this paper applies while its linear cosmological perturbations are allegedly stable. It remains to be seen whether the theory is viable with respect to static, spherically symmetric solutions and cosmological structure formation.

\acknowledgments
Thanks to Angelo Caravano and Marvin Lüben for many interesting discussions on the subject and to Mikica Kocic, Francesco Torsello, and an anonymous referee for comments on the manuscript.\\

\noindent \textbf{Note added:} Ref.~\cite{Caravano:2021aum} appeared at the same time as this paper. In the former paper, the authors analyze the constraints on the physical parameters from cosmological data and local observations for bimetric models with three parameters or less. Their results are consistent with ours, when comparable.\newpage

\appendix
\section{List of constraints}
\label{sec:ConstraintsList}
This section contains a complete list of the constraints discussed in the paper. 

\paragraph{Cosmological constraints.} In order for the matter density $\Omega_m$ to be positive in the early universe (hence a real-valued expansion rate),
\begin{equation}
1+2\alpha+\beta >0.
\end{equation}
To have a real-valued expansion rate in the final de Sitter phase,
\begin{equation}
\omegaleff >0.
\end{equation}
To avoid the Higuchi ghost, it is necessary and sufficient to impose,
\begin{equation}
\frac{\mfphat^2}{2 \omegaleff} > \frac{1}{2} \frac{3 \sec^2 \theta \sqrt{1+2\alpha+\beta}}{\sqrt{\beta}(3\tan^2 \theta [1+2\alpha+\beta]-\beta) + \sqrt{1+2\alpha+\beta}(1-\alpha + \beta -3 \tan^2 \theta [\alpha+\beta])},
\end{equation}
and,
\begin{subequations}
	\begin{alignat}{2}
	\beta &> -1-2\alpha,& \quad \alpha &> -1/2,\\
	\beta &> \frac{1}{12} \left(S_1 + 2\alpha \left[-6 + \frac{6}{1-3 \tan^2 \theta} + 3\alpha - S_1 \right] + \frac{3S_2}{1-3 \tan^2 \theta} -6 \right),& \quad \alpha &< -1/2,
	\end{alignat}
\end{subequations}
where,
\begin{subequations}
	\begin{align}
	S_1 &\equiv \sqrt{9(1+\alpha)^2 - 12 (1+2\alpha) \cos^2 \theta},\\
	S_2 &\equiv \sqrt{3 \sec^2 \theta (-1-2\alpha+3\alpha^2 +3 \tan^2 \theta (1+\alpha)^2)},
	\end{align}
\end{subequations}
We also noted that there is a constraint on the physical parameters if $\omegaleff > 1$ due to the requirement of a finite branch solution for $y$ with positive matter density at $z=0$. However, we did not work out the details of that constraint since $\omegaleff > 1$ is observationally excluded.

\paragraph{Local constraints.} To have a working Vainshtein screening mechanism that restores GR on local scales (e.g., in the solar system),
\begin{subequations}
	\label{eq:LocalConstr}
	\begin{align}
	\beta &> 1,\\
	\alpha &> - \sqrt{\beta},\\
	(\alpha + \sqrt{\beta})  \left(\alpha d_1+ d_2\right)&>0,
	\end{align}
\end{subequations}
where,
\begin{subequations}
	\begin{align}
	d_1 &\equiv 1 + 3 \tan^2 \theta -6 \sqrt{\beta} \sec^2 \theta + 3 \beta \sec^2 \theta,\\
	d_2 &\equiv -1 + 6\sqrt{\beta}(1+\beta) \sec^2 \theta - \beta (13 + 12 \tan^2 \theta),
	\end{align}
\end{subequations}

\paragraph{Other constraints.} To have the correct sign on the kinetic term for $\f$,
\begin{equation}
\theta \in \mathbb{R} \quad \Leftrightarrow \quad \kappa_g / \kappa_f >0.
\end{equation}
Linearly stable proportional solutions require,
\begin{equation}
\mfphat^2 >0.
\end{equation}

\section{The Stückelberg field}
\label{sec:muEq}
The ansatz for the metrics of a bidiagonal, static, spherically symmetric space-time reads,
\begin{equation}
	ds_g^2 = -A(r)^2 dt^2 + B^{-2}(r) dr^2 +r^2 d\Omega^2, \quad ds_f^2 = - \wt{A}^2(r) dt^2 + \frac{\left[\partial_r R(r)\right]^2}{\wt{B}^2(r)} dr^2 + R^2(r) d\Omega^2.
\end{equation}
The $R(r)$ function is a Stückelberg field in the sense that it is the area radius of $\f$ and hence encodes the coordinate transformation $r \to R$, that is from the radial coordinate being the area radius of $\g$ to the area radius of $\f$. In other words, $R$ can be used as the radial coordinate and the second metric then takes the same form as $\g$ in the ordinary $tr$-coordinates. The $\mu(r)$ function is related to the Stückelberg field $R(r)$ via (see Ref.~\cite{Enander:2015kda} for details),
\begin{equation}
	\mu(r) = R(r)/r-1,
\end{equation}
and hence we refer also to $\mu$ as the Stückelberg field. The seventh degree polynomial determining the function $\mu(r)$ is,
\begin{align}
\label{eq:muPoly}
	3 \sec^2 \theta \, \mu + 6 \sec^2 \theta \, (1- \alpha) \mu^2 +& \nonumber\\
	\frac{1}{3} \left[6 \sec^2 \theta \, \alpha^2 -2 (17+18 \tan^2 \theta)\alpha +4 \sec^2 \theta \, \beta +10 + 9 \tan^2 \theta \right] \mu^3+& \nonumber\\
	\frac{2}{3} \left[6 \sec^2 \theta \, \alpha^2 - (7+9 \tan^2 \theta)\alpha + 4 \sec^2 \theta \, \beta +1 \right] \mu^4 +& \nonumber\\
	\frac{1}{3} \left[2(1+3 \tan^2 \theta) \alpha^2 - \sec^2 \theta \, \beta^2 + 2(1+2 \tan^2 \theta) \beta - 4 \alpha \beta -2 \alpha \right] \mu^5 +& \nonumber\\
	-\frac{2}{3} \tan^2 \theta \, \beta^2 \mu^6 - \frac{1}{3} \tan^2 \theta \, \beta^2 \mu^7& \nonumber\\
	= - \lambda_g^2 \sec^2 \theta \, (1+\mu)^2 \left[\frac{2M(r)}{r^3} (1-\beta \mu^2) - \kappa_g P(r) (1 - 2 \alpha \mu + \beta \mu^2)\right],
\end{align} 
where $M(r)$ is the mass inside the radius $r$ and $P(r)$ is the pressure at $r$,
\begin{equation}
	M(r) = \kappa_g \int_{0}^{r} \rho(r) r^2 dr, \quad P(r) = {T^i}_i / 3 \; \mathrm{(summation \; implied)},
\end{equation}
and $\rho$ being the matter density $\rho = - {T^0}_0$. Solving \eqref{eq:muPoly} for $\mu$, the physical root is the one satisfying $\mu=-1/\sqrt{\beta}$ at $r=0$ and $\mu=0$ at $r=\infty$. The latter condition is implied by the requirement of an asymptotically flat solution. The physical root is guaranteed to exist if the conditions \eqref{eq:LocalConstr} are satisfied, see Ref. \cite{Enander:2015kda} for more details.

\section{Constraints from the dynamical Higuchi bound}
\label{sec:Higuchi}
The dynamical Higuchi bound can be written,
\begin{equation}
\frac{\mfphat^2}{2\omegaleff} > R(y;\theta,\alpha,\beta),
\end{equation}
where the ratio $R$ is,
\begin{subequations}
	\begin{alignat}{2}
	R &\equiv& &N/D,\\
	N &\equiv& &3 \sec^2 \theta \, y^3,\\
	D &\equiv& &1+2\alpha+\beta + 3 [\tan^2 \theta (1+2\alpha+\beta) - \beta] y^2 +\\ \nonumber
	& & & + 2[1-\alpha+\beta-3 \tan^2 \theta (\alpha+\beta)] y^3 + 3 \tan^2 \theta \, \beta y^4,
	\end{alignat}
\end{subequations}
see \eqref{eq:DynHig3}. A necessary condition for this equation to hold is that the denominator $D$ does not cross zero for any $y \in [0,1]$. One can show that this is the case precisely if,
\begin{subequations}
	\label{eq:DynHig4}
	\begin{alignat}{2}
	\label{eq:B1constr}
	\beta &> -1-2\alpha,& \quad \alpha &> -1/2,\\
	\label{eq:betaconstr}
	\beta &> \frac{1}{12} \left(S_1 + 2\alpha \left[-6 + \frac{6}{1-3 \tan^2 \theta} + 3\alpha - S_1 \right] + \frac{3S_2}{1-3 \tan^2 \theta} -6 \right),& \quad \alpha &< -1/2,
	\end{alignat}
\end{subequations}
with,
\begin{subequations}
	\begin{align}
	S_1 &\equiv \sqrt{9(1+\alpha)^2 - 12 (1+2\alpha) \cos^2 \theta},\\
	S_2 &\equiv \sqrt{3 \sec^2 \theta (-1-2\alpha+3\alpha^2 +3 \tan^2 \theta (1+\alpha)^2)}.
	\end{align}
\end{subequations}
Note that the first constraint \eqref{eq:B1constr} is just the requirement that $B_1$ is positive \eqref{eq:PosBeta1}. The right-hand side of \eqref{eq:betaconstr} depends on $\theta$ but only weakly; the most restrictive constraint is when $\theta=0$ and the least restrictive constraint is when $\theta = \pi/2$. The intermediate cases inerpolates between these two extremes.
\begin{subequations}
	\begin{alignat}{2}
	\beta|_{\theta=0} &> \frac{1}{6} (\alpha-1) \left(3 + 3\alpha - \sqrt{9\alpha^2 - 6\alpha - 3}\right),& \quad \alpha &< -1/2,\\
	\beta|_{\theta = \pi/2} &> \frac{1}{2}\left[(\alpha-2)\alpha -(1+\alpha)|1+\alpha| -1\right] ,& \quad \alpha &< -1/2.
	\end{alignat}
\end{subequations}
The next step is to find the maximum of $R$ on the interval $y\in[0,1]$. The necessary and sufficient condition for the dynamical Higuchi bound to be satisfied can then be written $\mfphat^2 /(2\omegaleff)>R_\mathrm{max}$. Differentiating,
\begin{equation}
\label{eq:dRdy}
\partial_y R(y;\theta,\alpha,\beta) = \left[f(y;\theta,\alpha,\beta)\right]^2 [1+2\alpha +\beta (1-y^2)],
\end{equation}
where $f(y;\theta,\alpha,\beta)$ is some real-valued function. So, the sign of the derivative is determined by the second factor. From this expression, we see that if $\alpha > -1/2$ then $\partial_y R >0$ for all $y \in [0,1]$ in which case $R$ obtains its maximum at $y=1$ and \eqref{eq:DynHig3} reduces to the ordinary Higuchi bound.

On the other hand, if $\alpha < -1/2$ and assuming $1 + 2 \alpha + \beta >0$, $\partial_y R$ switches sign from positive to negative at some point on the interval $y \in [0,1]$. In other words, $R$ has a maximum value in the interval $0<y<1$. Setting \eqref{eq:dRdy} to zero to find the maximum point,
\begin{equation}
y_\mathrm{max} = \sqrt{(1+ 2\alpha+\beta)/\beta}, \quad \alpha < -1/2. 
\end{equation}
Hence, the maximum value of $R$ is,
\begin{equation}
R_\mathrm{max} = \frac{1}{2} \frac{3 \sec^2 \theta \sqrt{1+2\alpha+\beta}}{\sqrt{\beta}(3\tan^2 \theta [1+2\alpha+\beta]-\beta) + \sqrt{1+2\alpha+\beta}(1-\alpha + \beta -3 \tan^2 \theta [\alpha+\beta])},
\end{equation}
and the dynamical Higuchi constraint reads,
\begin{equation}
\label{eq:DynHig3negAlph}
\frac{\mfphat^2}{2 \omegaleff} > R_\mathrm{max}, \quad \alpha < -1/2.
\end{equation}
The constraint depends weakly on $\theta$ and is most restrictive in the $\theta \to 0$ limit,
\begin{subequations}
	\begin{alignat}{2}
	R_\mathrm{max}|_{\theta=0} &= \frac{3}{2} \left(1 - \alpha + \beta \left[1 - \sqrt{\frac{\beta}{1+2\alpha+\beta}}\right]\right)^{-1},& \quad \alpha &< -1/2,\\
	R_\mathrm{max}|_{\theta = \pi / 2} &= \frac{1}{2} \left(-(\alpha+\beta) + \sqrt{\beta (1+ 2\alpha+\beta)}\right)^{-1},& \quad \alpha &< -1/2,
	\end{alignat}
\end{subequations}
see Fig. \ref{fig:dynhig3} for exclusion plots based on the theoretical constraints from bimetric cosmology.

\section{Expansions}
\label{sec:dSexp}
\subsection{Around the final de Sitter point}
Expanding $E^2$ around $\Omega_m= 0$ (i.e., the asymptotic future de Sitter state),
\begin{equation}
\label{eq:Eexp_full}
E^2 = \omegaleff + \frac{\kappa_g^\mathrm{eff}}{\kappa_g} \Omega_m + c_2 \Omega_m^2 + c_3 \Omega_m^3 + c_4 \Omega_m^4  + \mathcal{O}(\Omega_m^5),
\end{equation}
where
\begin{subequations}
	\label{eq:Ecoeffs}
	\begin{align}
		\kappa_g^\mathrm{eff} &= \kappa_g \left(1 - \sin^2 \theta \frac{\mfphat^2 }{\mfphat^2  - 2 \omegaleff}\right),\\
		c_2 &= (1 + 2 \alpha) \sin^2 \theta \frac{\mfphat^2 \omegaleff}{(\mfphat^2 - 2 \omegaleff)^3},\\
		c_3 &= \frac{\mfphat^2}{6 (\mfphat^2 - 2 \omegaleff)^5} \sin^2 \theta \left[\mfphat^4 (1+2\alpha) - 4 \omegaleff^2 (3 + 6\alpha + 2 \beta) \right.\nonumber \\
		&\left. \cos (2\theta) \mfphat^2 (\mfphat^2 - 2 \omegaleff)(1+2\alpha) - 2 \mfphat^2 \omegaleff (1 + 4\alpha(2+3\alpha) -2\beta)\right],\\
		c_4 &= \frac{\sin^2(2\theta)}{6 (\mfphat^2 - 2 \omegaleff)^7} \Big[ \mfphat^8 (1 + \beta - \alpha (1+6\alpha)) +3 \sec^2 \theta \, \mfphat^2 \omegaleff^3 (5 + 10\alpha + 4\beta)  \nonumber \\
		&+ 2 \mfphat^4 \omegaleff^2 (7 + \beta - \tan^2 \theta \, \beta + 5 \tan^2 \theta \, \alpha (3 + 6\alpha +2\beta) + \alpha (33 + 38\alpha + 10\beta)) \nonumber \\
		&+ \mfphat^6 \omegaleff (-9 + (19+15\tan^2 \theta)\alpha^2 + 30 \sec^2 \theta \, \alpha^3 -2(3+\tan^2 \theta )\beta - 2\alpha (8 + 5 \sec^2 \theta \beta))  \Big].
	\end{align}
\end{subequations}
We also calculated the $\Omega_m^5$ term but due to its length, we do not display it here. A difference between the $\mfphat \to \infty$ limit and the $\theta \to 0$ limit is that in the former, the transition from the final $\Lambda$CDM phase to the bimetric phase is pushed backwards in time while in the latter, the bimetric modification is suppressed with the same factor at all redshifts. This can be seen as the different terms in the expansion \eqref{eq:Ecoeffs} are suppressed with different leading order powers of $1/\mfphat$ in the large graviton mass limit. For example, the $\Omega_m^2$ term is suppressed by $1/\mfphat^4$,
\begin{equation}
\left[\frac{1}{\mfphat^4} \sin^2 \theta \, \omegaleff (1+2\alpha)  + \mathcal{O}\left(\frac{1}{\mfphat^6}\right)\right] \Omega_m^2,
\end{equation}
while the $\Omega_m^4$ term is suppressed by $1/\mfphat^6$,
\begin{equation}
\left[\frac{1}{\mfphat^6} \frac{1}{6} \sin^2 (2\theta) (1 - \alpha(1+6\alpha) + \beta)  + \mathcal{O}\left(\frac{1}{\mfphat^8}\right)\right] \Omega_m^4.
\end{equation}
Doing the same analysis in the $\theta \to 0$ limit, we find that all terms up to order $\Omega_m^5$ are suppressed with $\tan^2 \theta$, so the bimetric modification is suppressed by the same factor for all redshifts in the $\theta \to 0$ limit.

\begin{figure}[t]
	\centering
	\includegraphics[width=0.9\linewidth]{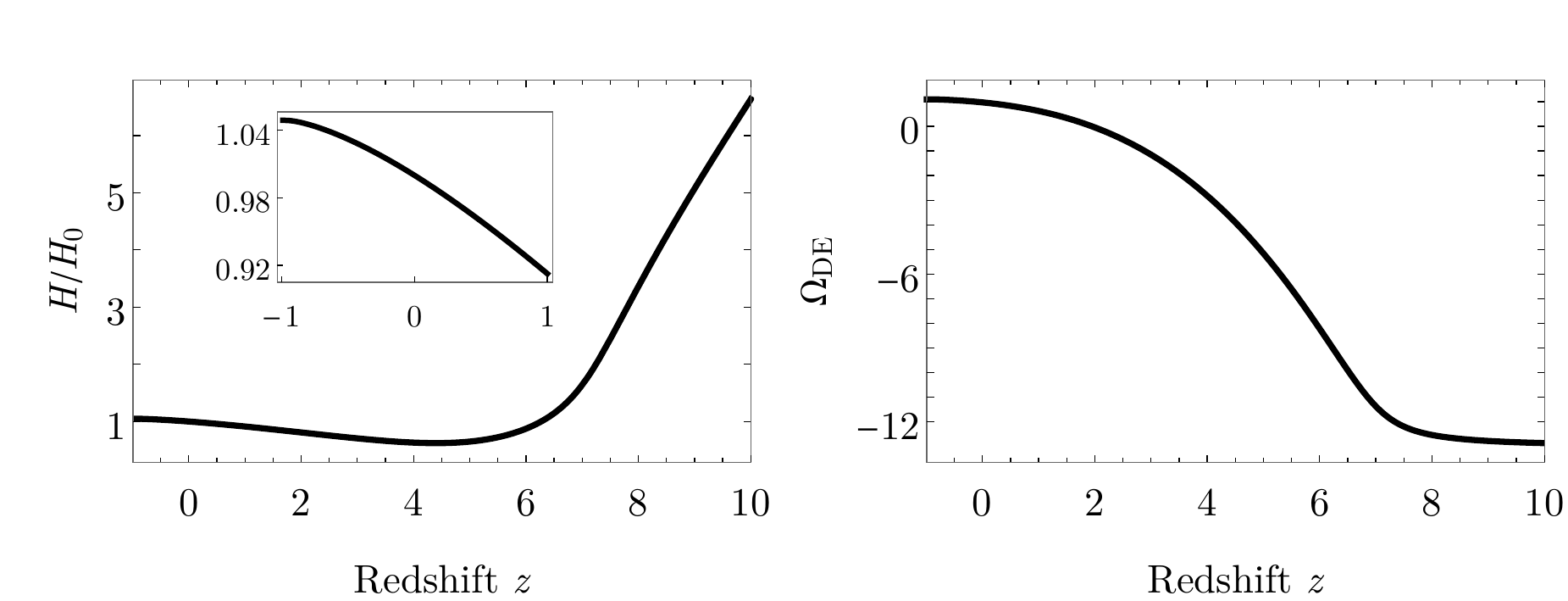}
	\caption{Expansion rate $E = H/H_0$ (left panel) and dark energy density $\omegade$ (right panel) as functions of redshift. Here we have set $(\theta,\mfphat,\omegaleff,\alpha,\beta)=(84^{\circ},1.5,1.1,10,10)$ so that the effective gravitational constant is negative in the late universe, $\kgeff = -44 \kappa_g$. As shown in the left panel the expansion rate increases in the future ($z \to -1$) in this case. This allows for an effective cosmological constant greater than one, here $\omegaleff = 1.1$.}
	\label{fig:erepulsivegrav}
\end{figure}

The factor $\kgeff$ in front of $\Omega_m$ in the expansion can be interpreted as the effective (background) gravitational constant of the late-time universe. If $\omegaleff=0$ (or $\mfphat^2 \gg \omegaleff$), the effective gravitational constant \eqref{eq:kgeff} agrees with the $\kgeffm$ for the local solutions far outside the Compton wavelength \eqref{eq:kappagMassive}, as expected. Due to the Higuchi bound \eqref{eq:Higuchi1}, this gravitational constant is always less than $\kappa_g$, so the effective gravitational force is stronger in the early universe. However, the gravitational constant also depends on the length scale. As an example, for a $B_1$-model (i.e., with all other $B$-parameters zero), $\kgeff = \kappa_g / 2$. For large values of $\mfphat^2/2 \omegaleff$, the effective gravitational constant is negative, hence gravity is repulsive on large scales in the late universe. This happens precisely when $E' > 0$ meaning that instead of continuing to decrease, $E$ increases with time and the effective cosmological constant can be greater than one, $\omegaleff >1$, see Fig.~\ref{fig:erepulsivegrav} for an example.

\subsection{$\mfphat \to \infty$ limit}
\label{sec:Expm}
To calculate the expansion rate $E$, we first solve \eqref{eq:BRFriedm} and \eqref{eq:yPoly} evaluated today (i.e., at $z=0$) to obtain $y_0$ and $\omegamnot$. (Recall that $E_0=1$.) Then we solve \eqref{eq:BRFriedm} and \eqref{eq:yPoly} for $y$ and $E$ at each redshift we are interested in. When $\mfphat \gg 1$, one can expand $y_0$, $\omegamnot$, $y$, and $E$ as Taylor series around $1/\mfphat=0$ and solve the equations order by order. The result is,
\begin{subequations}
	\begin{alignat}{2}
		y_0 &= \sum_{n=0}^{\infty} \mathfrak{a}_n \left(\frac{1}{\mfphat^2}\right)^n,& \quad \omegamnot &= \sum_{n=0}^{\infty} \mathfrak{b}_n \left(\frac{1}{\mfphat^2}\right)^n,\\
		y &= \sum_{n=0}^{\infty} \mathfrak{c}_n \left(\frac{1}{\mfphat^2}\right)^n,& \quad E^2 &= \sum_{n=0}^{\infty} \mathfrak{d}_n \left(\frac{1}{\mfphat^2}\right)^n.
	\end{alignat}
\end{subequations}
The first coefficients are given by,
\begin{subequations}
	\label{eq:ExpCoeffm}
	\begin{align}
		\mathfrak{a}_0 &= 1,\\
		\mathfrak{a}_1 &= -\sec^2 \theta (1-\omegaleff),\\
		\mathfrak{b}_0 &= \sec^2 \theta (1-\omegaleff),\\
		\mathfrak{b}_1 &= 2 \tan^2 \theta \, \sec^2 \theta \, \omegaleff (1-\omegaleff),\\
		\mathfrak{c}_0 &= 1,\\
		\mathfrak{c}_1 &= - \sec^2 \theta (1-\omegaleff) \omhat(z),\\
		\mathfrak{d}_0 &= \omegaleff + (1-\omegaleff) \omhat(z),\\
		\mathfrak{d}_1 &= 0,\\
		\mathfrak{d}_2 &= - \frac{1}{3} \tan^2 \theta \, \sec^2 \theta \, (1+2\alpha) (1-\omegaleff)^2 \omhat(z) \left(1 - \omhat(z)\right) \left[1 + 2\omegaleff + (1-\omegaleff) \omhat(z)\right].
	\end{align}
\end{subequations}
Here we have introduced $\omhat(z)$ as a shorthand for,
\begin{equation}
	\omhat(z) \equiv \Omega_m(z) / \omegamnot = (1+z)^{3(1+w_m)}.
\end{equation}
From \eqref{eq:ExpCoeffm}, we see that in the infinite graviton mass limit, $y \to 1$ and $\omegamnot \to \sec^2 \theta (1-\omegaleff)$ and the expansion rate $E$ approaches a $\Lambda$CDM model with cosmological constant $\omegaleff$ and matter density today $1-\omegaleff$. We calculated $\mathfrak{a}_n$ up to $n=5$, $\mathfrak{b}_n$ up to $n=4$, $c_n$ up to $n=4$, and $c_n$ up to $n=3$. Here, we do not write out the higher order coefficients, due to their length.

\subsection{$\alpha \to \infty$ limit}
\label{sec:ExpAlpha}
In this case, $y_0$, $\omegamnot$, $y$, and $E$ can be expanded as a Puiseux series in positive powers of $\alpha^{-1/2}$ and the equations are solved order by order in this small parameter. The result is,
\begin{subequations}
	\begin{alignat}{2}
	y_0 &= \sum_{n=0}^{\infty} \mathfrak{a}_n \left(\frac{1}{\alpha^{1/2}}\right)^n,& \quad \omegamnot &= \sum_{n=0}^{\infty} \mathfrak{b}_n \left(\frac{1}{\alpha^{1/2}}\right)^n,\\
	y &= \sum_{n=0}^{\infty} \mathfrak{c}_n \left(\frac{1}{\alpha^{1/2}}\right)^n,& \quad E^2 &= \sum_{n=0}^{\infty} \mathfrak{d}_n \left(\frac{1}{\alpha^{1/2}}\right)^n.
	\end{alignat}
\end{subequations}
The first coefficients read,
\begin{subequations}
	\begin{align}
		\mathfrak{a}_0 &= 1,\\
		\mathfrak{a}_1 &= \pm \frac{1}{\mfphat} \sec \theta \sqrt{(1-\omegaleff)},\\
		\mathfrak{b}_0 &= \sec^2 \theta (1-\omegaleff),\\
		\mathfrak{b}_1 &= \pm 2 \tan^2 \theta \, \sec \theta \frac{ (1-4\omegaleff) \sqrt{(1-\omegaleff)}}{3 \mfphat},\\
		\mathfrak{c}_0 &= 1,\\
		\mathfrak{c}_1 &= \pm \frac{1}{\mfphat} \sec \theta \sqrt{(1-\omegaleff) \omhat(z)},\\
		\mathfrak{d}_0 &= \omegaleff + (1-\omegaleff) \omhat(z),\\
		\mathfrak{d}_1 &= \pm \frac{2 \sin^2 \theta}{3 \mfphat \cos \theta} \sqrt{1-\omegaleff} \left[(1-4\omegaleff) \omhat(z) + 3\omegaleff \sqrt{\omhat(z)} - (1-\omegaleff) \omhat^{3/2}(z)\right].
	\end{align}
\end{subequations}
Recall that $\omhat(z) \equiv (1+z)^{3(1+w_m)}$. In some coefficients there is a choice of plus/minus sign. The plus sign is for infinite branch solutions and the minus sign for finite branch solutions. As in the infinite graviton mass limit, $y \to 1$ and $\omegamnot \to \sec^2 \theta (1-\omegaleff)$ as $\alpha \to \infty$ and the expansion rate $E$ approaches a $\Lambda$CDM model with cosmological constant $\omegaleff$ and matter density today $1-\omegaleff$.

\subsection{$\beta \to \infty$ limit}
\label{sec:ExpBeta}
Here, we expand $y_0$, $\omegamnot$, $y$, and $E$ as a Puiseux series in positive powers of $\beta^{-1/3}$ and the equations are solved order by order in this small parameter. The result is,
\begin{subequations}
	\begin{alignat}{2}
	y_0 &= \sum_{n=0}^{\infty} \mathfrak{a}_n \left(\frac{1}{\beta^{1/3}}\right)^n,& \quad \omegamnot &= \sum_{n=0}^{\infty} \mathfrak{b}_n \left(\frac{1}{\beta^{1/3}}\right)^n,\\
	y &= \sum_{n=0}^{\infty} \mathfrak{c}_n \left(\frac{1}{\beta^{1/3}}\right)^n,& \quad E^2 &= \sum_{n=0}^{\infty} \mathfrak{d}_n \left(\frac{1}{\beta^{1/3}}\right)^n.
	\end{alignat}
\end{subequations}
The first coefficients are,
\begin{subequations}
	\begin{align}
		\mathfrak{a}_0 &= 1,\\
		\mathfrak{a}_1 &= - \left(\frac{3\sec^2 \theta (1-\omegaleff)}{\mfphat^2}\right)^{1/3},\\
		\mathfrak{b}_0 &= \sec^2 \theta (1-\omegaleff),\\
		\mathfrak{b}_1 &= - \tan^2 \theta (1-3\omegaleff) \left(\frac{3 \sec^2 \theta(1-\omegaleff)}{\mfphat^2}\right)^{1/3},\\
		\mathfrak{c}_0 &= 1,\\
		\mathfrak{c}_1 &= - \left(\frac{3\sec^2 \theta(1-\omegaleff)}{\mfphat^2} \omhat(z)\right)^{1/3},\\
		\mathfrak{d}_0 &= \omegaleff + (1-\omegaleff)\omhat(z),\\
		\mathfrak{d}_1 &= \tan^2 \theta \left[\frac{3 \cos^4 \theta (1-\omegaleff)}{\mfphat^2} \omhat(z)\right]^{1/3} \left[(1-\omegaleff) \omhat(z) - (1-3\omegaleff) \omhat^{2/3}(z) -2\omegaleff\right]. 
	\end{align}
\end{subequations}
As in the $\mfphat \to \infty$ and $\alpha \to \infty$ limits, $y \to 1$ and $\omegamnot \to \sec^2 \theta (1-\omegaleff)$ as $\beta \to \infty$. Also, the expansion rate $E$ approaches a $\Lambda$CDM model with cosmological constant $\omegaleff$ and matter density today $1-\omegaleff$.

\section{Avoiding the Big Rip}
\label{sec:BigRip}
In general, the bimetric fluid has a phantom equation of state. Here, we show that it approaches $w_\mathrm{DE}=-1$ fast enough in the late universe
so that the Big Rip is avoided. In the limits $\mfphat, \alpha,\beta \to \infty$, the background cosmology approaches $\Lambda$CDM and the Big Rip is avoided. If we are not in these limits, we can expand the expansion rate $E$ in the late universe around $\Omega_m=0$ as in \eqref{eq:Eexp_full}. Using $E = H / H_0 = (da/dt)/H_0$, we can write, to first order in $\Omega_m$,
\begin{equation}
	\left(\frac{da/dt}{H_0 a}\right)^2 = \omegaleff + \frac{\kgeff}{\kappa_g} \Omega_m.
\end{equation}
Solving for $dt$ and integrating,
\begin{equation}
	H_0 \Delta t = \int \frac{da}{a} \left[\omegaleff + \frac{\kgeff}{\kappa_g} \Omega_m\right]^{-1/2},
\end{equation}
where $\Delta t$ is the time to Big Rip. Changing integration variable from $a$ to $\Omega_m$, using \eqref{eq:OmegaMcons}, the integral evaluates to,
\begin{equation}
	H_0 \Delta t = \frac{2}{3 (1+w_m)\sqrt{\omegaleff}} \mathrm{arctanh} \left[\sqrt{1 + \frac{\kgeff}{\kappa_g \omegaleff} \Omega_m}\right],
\end{equation}
which is positive if $w_m > -1$ and diverges as $\Omega_m \to 0$ (i.e., $a \to \infty$). Thus, it takes infinite time to reach an infinite scale factor, that is, there is no Big Rip.

\bibliographystyle{JHEP}
\bibliography{biblio}

\end{document}